\documentclass{article}
\usepackage[utf8]{inputenc}
\usepackage{amsmath,epsfig,epsf,psfrag}
\usepackage{amsfonts}
\usepackage{amssymb,bm}
\usepackage{graphicx}
\usepackage{natbib}
\usepackage{booktabs}

\begin{document}
\title{Direct Sampling of Bayesian Thin-Plate Splines for Spatial Smoothing}
\author{Gentry White\footnote{gentry.white@qut.edu.au}\\Mathematical and Statistical Sciences School\\ Queensland University of Technology\\ Brisbane, QLD, 4001, Australia\vspace{3mm}\\ 
Dongchu Sun\\Department of Statistics\\ University of Missouri\\ Columbia,MO 62511\vspace{3mm}\\
Paul Speckman\\Department of Statistics\\University of Missouri\\ Columbia,MO 62511}

\setlength{\textheight}{575pt}

\maketitle
\begin{abstract}
Radial basis functions are a common mathematical tool used to construct a smooth interpolating function from a set of data points.  A spatial prior based on thin-plate spline radial basis functions can be easily implemented resulting in a posterior that can be sampled directly using Monte Carlo integration, avoiding the computational burden and potential inefficiency of an Monte Carlo Markov Chain (MCMC) sampling scheme.  The derivation of the prior and sampling scheme are demonstrated.  
\end{abstract}

\section{Introduction}
Noisy or incomplete spatial data occurs in many contexts, and the detection of trends in or the identification of clusters or other anomalies is often the central question of interest in exploratory
data analysis.  In this context, the goal is ``smoothing'', i.e. removing noise from the observed data while preserving the underlying spatial patterns.
Bayesian methods for smoothing are common, offering benefits in terms of model specification, and allowing inference directly from the posterior distribution rather than relying on the asymptotic approximations of classical inference \citep{Banerjee:etal:2015}.  The downside of Bayesian methods is that their computational burden can be substantially higher than classical approaches.  Inference for Bayesian models is based on a using the posterior distribution which can typically not be obtained analytically and requires using methods to draw samples from the posterior distribution.  Ideally these samples are drawn directly from the joint posterior of all the model parameters, but in practice this is typically not possible, necessitating the use of Monte Carlo Markov Chain (MCMC) schemes which produce correlated samples reducing effective sample size, and can require a substantial ``burn-in'' to ensure that the samples are drawn from the target distribution \citep{Walker:etal:2011}.   In some cases, Bayesian models, including spatial smoothers can be evaluated using integrated-nested Laplace approximations (INLA) methods \citep{Rue:etal:2009}, but these methods rely on approximations to the posterior distributions. 

This paper presents a prior distribution for spatial effects derived from thin-plate spline radial basis functions and a sampling scheme that allows for direct sampling from the joint posterior via Monte Carlo integration. 
This avoids the use of a time-consuming MCMC scheme
while drawing samples directly from the actual posterior distribution in roughly the same time required to draw samples from the INLA approximation of the posterior distribution.

Statistical methods for the analysis of spatial data date trace their beginnings to the 1960s and the advent of Geo-statistical models   \citep{Matheron:1963}, and lattice models in the 1970s \citep{Besag:1974}, which form the basis of many modern methods of analysis for spatial data.  
While these methods are inherently statistical in their foundations, radial basis functions arise in applied  and computational mathematics from the goal of creating a linear interpolation of observed data in order to approximate a complex or intractable function \cite{Powell:1977,Broomhead:Lowe:1988}. The connection between this and the desire to create a smooth image of spatially varying data may seem esoteric, but noting that radial basis functions create a smooth interpolating function based on observed data, and spatial smoothing seeks a smooth spatially varying function that approximates the observed data while minimising noise, the similarities become clearer.  

Multiple researchers have explored the connection between radial basis functions, specifically thin-plate splines, and smoothing splines, including for spatial smoothing. \cite{Wahba:1990} and \cite{Green:Silver:1994} provide a thorough technical and historical coverage of smoothing splines, and the use of radial basis function, specifically thin-plate splines, as spatial smoothers can be found in \cite{Wahba:etal:1995} and \cite{vanderLinde:etal:1995}.  Comparison between the use of thin-plate splines, other non-parametric
smoothing functions, and more traditional geo-statistical techniques including kriging are made in \cite{Laslett:1994}, \cite{Hutchinson:Gessler:1994}, \cite{Laslett:McBrat:1990}, and \cite{Nychka:2000}, which provides several examples of applications. 
Using the Bayesian framework suggested in \cite{Wahba:1978}, \cite{Wahba:1983}, \cite{Kimmeldorf:Wahba:1970} and \cite{Kimmeldorf:Wahba:1971}, and results from \cite{White:2006} a spatial prior is derived based on thin-plate splines which provides a computationally efficient implementation that doesn't require the use of Monte Carlo Markov Chain (MCMC) methods for evaluation, and a straightforward interpretation of results.  

In Section \ref{methods} of this paper the thin-plate spline smoothing solution is derived as a solution to a Bayesian hierarchical model.  In Section \ref{comp} this model is further refined and the prior distributions are derived to allow the implementation of a computational scheme that allows for drawing samples directly from the posterior distribution, rather than relying on an MCMC scheme.  In Section \ref{examples} the computational and smoothing results of this prior are demonstrated using several example datasets from a variety of applications.  In Section \ref{disc} the results are discussed in their contexts. 
\section{Methods and Computation}
\label{methods}
Radial basis functions are a common mathematical tool used to create a smooth interpolating surface as a means of approximating a function from observed data.  In general terms given the observed pairs $(y,\bm{x})$ the relationship
\begin{eqnarray}
y(\bm{x})=\sum_{i=1}^n\alpha_i\psi(\|\bm{x}-\bm{x}_i\|),
\end{eqnarray}
can be written for an appropriate set of basis functions $\phi(\cdot)$ computed for the Euclidean distance between observations. The specific properties of the basis functions create a system of linear equations that can be shown to have a unique solution for the weights $\alpha_i$.  
In the context of spatial data, this corresponds to fitting a smooth interpolating surface over a set of observed values at given locations; assuming there is no noise in the observations.  In practice when smoothing spatial data it is assumed that there is an underlying process (a smooth function) describing the spatial variation in data and that the observations contain noise.  The radial basis function model simply interpolates observed data without considering noisy observations, so using radial basis functions to construct a spatial smoother requires modifications to allow for noisy observations.  This is done by modelling the observed data with a hierarchical or mixed-effects model where the data are assumed to follow a distribution with a mean that is the function of a random spatial effect whose prior distribution incorporates radial basis functions in its covariance structure as a means of describing spatial variation.  
The derivation here is presented to illustrate the derivation of a spatial prior based on thin-plate splines basis functions, for a broader and more detailed treatment of splines and and their statistical application see \cite{Wahba:1990}, \cite{Gu:2002} or \cite{Nychka:2000}

\subsection{Thin Plate Splines and Their Solution as a Smoother}
A thin-plate spline smoother over a $d$-dimensioned surface can be defined as the solution that minimizes the penalized sum 
of squares
\begin{eqnarray}
\label{S}
S_{\eta}(f) &=& \frac{1}{n}\sum^{n}_{i}W_i(y_i - f(\bm{x}_i))^2
+\eta J_m(f)
\end{eqnarray}
where $\eta>0$ and the penalty term
\begin{eqnarray}
\label{Jm}
J_m(f) &=&
\int_{\mathbb{R}^d}\sum\frac{m!}{\alpha_1!\dots\alpha_d!}
\left(\frac{\partial^mf}{\partial x_1^{\alpha_1}\dots \partial
x_d^{\alpha_d}}\right)^2d\bm{x}.
\end{eqnarray}

The sum in the integrand is taken over all the non-negative integer vectors
$\bm{\alpha}=(\alpha_1,\dots,\alpha_d)'$ such that $\sum \alpha_1+\dots +\alpha_d =
 m$, and $2m > d$. In the 
case of spatial data where $d=2$, $m = 2$.
\cite{Matheron:1973} and \cite{Duchon:1977} show that the solution belongs to the finite dimensional space
\begin{equation}
\label{f1}
f(\bm{x}) = \sum^t_{j = 1}\phi_j(\bm{x})\beta_j + \sum^n_{i =
1}\psi_i(\bm{x})\gamma_i,
\end{equation}
where $(\phi_1,\dots\phi_t)$ is a set of $t$ functions that span the space of 
all $d$-dimensioned
polynomials of degree less than $m$, and $(\psi_1\dots,\phi_n)$ is a set of $n$ thin-plate splines 
radial basis functions defined as
\begin{eqnarray}
\psi_i(\bm{x})& =&  
\left\{ \begin{array}{l@{\quad \quad}l}
a_{md}\|\bm{x}-\bm{x}_i\|^{(2m-d)}\log\|\bm{x}-\bm{x}_i\|,& \mbox{if $d$ is even} \\
a_{md}\|\bm{x}-\bm{x}_i\|^{(2m-d)},& \mbox{if $d$ is odd,}
\end{array} \right.
\end{eqnarray}
where $a_{md}$ are arbitrary constants.
  
In matrix notation write $\bm{T} = (T_{ij})$ and $\bm{K} = (K_{ij})$,
where $T_{ij} = \phi_j(\bm{x}_i)$ and $K_{ij} = \psi_i(\bm{x}_j)$. 
Then (\ref{f1}) is expressed as
\begin{eqnarray}
\label{f2}
\left(\begin{array}{c}
f(\bm{x}_1)\\
\vdots\\
f(\bm{x}_n)
\end{array}\right)& =& \bm{T}\bm{\beta} + \bm{K}\bm{\gamma}.
\end{eqnarray}  
 \cite{Meinguet:1979} and \cite{Duchon:1977} also show that  
 equation (\ref{Jm}) can be written as 
\begin{eqnarray}
\label{Jm2}
J_m(f(\bm{x})) &=& \bm{\gamma}'\bm{K}\bm{\gamma}.
\end{eqnarray}
Subject to the constraint that $\bm{T}'\bm{\gamma} = 0$, the minimization
problem (\ref{S}) becomes a constrained minimization problem with
objective function 
\begin{eqnarray}
\label{S2}
S_{\eta}(f(\bm{x}))&=&
(\bm{y}-\bm{T}\bm{\beta}-\bm{K}\bm{\gamma})'\bm{W} (\bm{y}-\bm{T}\bm{\beta}-\bm{K}\bm{\gamma})
+\eta\bm{\gamma}'\bm{K}\bm{\gamma}.
\end{eqnarray}
The problem is simplified by removing the explicit constraint that $\bm{T}\bm{\gamma}=0$ and making it implicit in the characterisation of the objective function.  This is accomplished as suggested in \cite{Wahba:1990}.  Let $\bm{T}\bm{T}' = \bm{F}\bm{\lambda}\bm{F}'$ be the
spectral decomposition, where $\bm{F}$ is the matrix of eigenvectors and , $\bm{\lambda}$ is the 
diagonal matrix of eigenvalues.
Let $\bm{F}=[\bm{F}_1,\bm{F}_2]$, where $\bm{F}_1$ is the $n\times t$ matrix of
vectors spanning the column space of $\bm{T}$. Noting that $\bm{T}'\bm{\gamma} = 0$ if
and only if $\bm{\gamma} = \bm{F}_2\bm{\lambda}$ for some $\bm{\lambda}$,
 the minimization problem in (\ref{S2}) can be written as 
\begin{equation}
\label{S3}
\min_{\bm{\beta},\bm{\lambda}}\:(\bm{y}-\bm{T}\bm{\beta}-\bm{K}\bm{F}_2\bm{\lambda})'\bm{W}
(\bm{y}-\bm{T}\bm{\beta}-\bm{K}\bm{F}_2\bm{\lambda})
+\eta\bm{\lambda}'\bm{F}_2'\bm{K}\bm{F}_2\bm{\lambda}.
\end{equation}
If we define the following matrices and vector
\begin{eqnarray*}
\bm{G}  = \left[\bm{T},\bm{K}\bm{F}_2\right]_{n \times n}, \:\:\:
\bm{H}  =  \left[\begin{array}{cc}0 & 0\\0 & \bm{F}_2'\bm{K}\bm{F}_2\end{array}
\right]_{n\times n},\:\:\:
\bm{\omega}  =  \left(\begin{array}{c}\bm{\beta} \\ \bm{\lambda}\end{array}\right),
\end{eqnarray*}
then (\ref{S3}) can be written as 
\begin{equation}
\label{S4}
\min_{\bm{\omega}}\:(\bm{y}-\bm{G\omega})'\bm{W}(\bm{y}-\bm{G\omega})
+\eta\bm{\omega}'\bm{H}\bm{\omega}.
\end{equation}
Note that if $\eta=\delta_0/\delta_1$, the objective function in \eqref{S4} is proportional to the log-posterior density of $\bm{\omega}$, given the prior 
\begin{equation}
\label{S4a}
\pi(\bm{\omega}|\eta)\propto\exp\left(-\delta_1\bm{\omega}^{'}\bm{H\omega}\right).
\end{equation}
and Gaussian likelihood for $\bm{y}$
\begin{eqnarray}
\label{eq:loglikey}
\bm{y}|\bm{\omega},\delta_0\sim N(\bm{G\omega},\delta_0\bm{I}).
\end{eqnarray}
 Under the Bayesian penalised splines problem \cite{Lang:Brezger:2001} with the set of basis functions $\bm{G}$ and the parameters or weights $\bm{\omega}$ this equates to prior on $\bm{\omega}$ which penalises roughness or model complexity, via the matrix $\bm{M}$ and the parameter $\eta$. 

The optimisation problem in \eqref{S4} can be re-written as a non-parametric (or semi-parametric) optimisation by letting $\bm{\nu}  =  \bm{G\omega}$ and $\bm{M} = (\bm{G}^{-1})^{'}\bm{H}\bm{G}^{-1}$ and re-writing (\ref{S4}) as
\begin{equation}
\label{S5}
\min_{\bm{\nu}}\:(\bm{y}-\bm{\nu})'\bm{W}(\bm{y}-\bm{\nu})+
\eta\bm{\nu}'\bm{M\nu}.
\end{equation}
Which, given $\eta$, has the smoothing solution 
\begin{eqnarray}
\label{S6}
\hat{\bm{\nu}}&=&
(\bm{W}+\eta\bm{M})^{-1}\bm{y}.
\end{eqnarray}   

\subsection{Thin-Plate Splines Prior}

The minimization problem in (\ref{S5}) has a Bayesian interpretation, first
suggested by \cite{Kimmeldorf:Wahba:1971} and \cite{Wahba:1978}. 
Suppose $\bm{y}$ follows a normal distribution
\begin{eqnarray}
\label{like}
\bm{y}&\sim& N(\bm{\nu},\delta_0\bm{W}^{-1}).
\end{eqnarray}
Define the prior of $\bm{\nu}$ as a partially improper prior with density function
\begin{equation}
\label{omega}
\left[\bm{\nu}\mid\delta_1\right]\propto\ \delta_1^{-{r}/2}
\exp\left(-\frac{1}{2\delta_1}\bm{\nu}'\bm{M\nu}\right),
\end{equation}
where $r$ is the rank of the matrix $\bm{M}$,(see \cite{Speckman:Sun:2003}).
With this prior, the log-posterior of $\bm{\nu}$ given $\bm{y}$, $\delta_0$ and
$\delta_1$ is (up to an additive constant)
\begin{equation}
\label{loglike}
-\frac{1}{2\delta_0} (\bm{y}-\bm{\nu})'\bm{W}({\bm{y}}-\bm{\nu})
-\frac{1}{2\delta_1}\bm{\nu}'\bm{M\nu}.
\end{equation}
Making the substitution $\delta_1=\delta_0/\eta$ \cite{Nychka:2000}, the resulting conditional posterior distribution of $\bm{\nu}$ is
\begin{equation}
\label{fcnu}
\bm{\nu}|\delta_0,\eta,\bm{y}\sim N\left((\bm{W}+\eta\bm{M})^{-1}\bm{y},\frac{1}{\delta_0}(\bm{W}+\eta\bm{M})^{-1}\right).
\end{equation}
Substituting $\delta_1=\delta_0/\eta$, makes the posterior expectation 
of $\bm{\nu}$ in \eqref{fcnu} equivalent to the smoothing solution 
\eqref{S6}.

In order to complete the hierarchical model the prior distributions for $\delta_0$ are $\delta_1$, are needed, e.g.
\begin{eqnarray}
\label{delta}
\pi(\delta_l) &\propto&
\frac{1}{\delta_0^{a_l+1}}\exp\left(-\frac{b_l}{\delta_0}\right),\:l=0,1.
\end{eqnarray}
The resulting conditional posterior distributions are
\begin{eqnarray}
\label{fc0}
\pi(\delta_0|\bm{\nu},\bm{y})&\sim& \mbox{InvGamma}\left(a_0+\frac{n}{2},
b_0+\frac{(\bm{y}-\bm{\nu})^T(\bm{y}-\bm{\nu})}{2}\right)\\
\label{fc1}
\pi(\delta_1|\bm{\nu},\bm{y})&\sim& \mbox{InvGamma}\left(a_1+\frac{n-3}{2},b_1+\frac{\bm{\nu}^T\bm{M\nu}}{2}\right).
\end{eqnarray}
Given the likelihood
\begin{eqnarray}
\label{likelihood}
f(\bm{y}|\bm{\nu},\delta_0)&=&\frac{1}{\left(2\pi\delta_0\right)^{n/2}}\exp\left[-\frac{1}{2\delta_0}\left(\bm{y}-\bm{\nu}\right)^T
\left(\bm{y}-\bm{\nu}\right)\right]
\end{eqnarray}
and the set of conditional distributions \eqref{fcnu}, \eqref{fc0}, and \eqref{fc1} the posterior distributions for the parameters $\bm{\nu}$,$\delta_0$ and $\delta_1$ is 
\begin{eqnarray}
\pi\left(\bm{\nu},\delta_0,\delta_1|\bm{y}\right)&\propto&
f(\bm{y}|\bm{\nu},\delta_0)\pi\left(\bm{\nu}|\eta,\delta_1\right)\pi\left(\delta_0\right)
\pi\left(\delta_1\right)
\end{eqnarray}
which has no closed form but can be evaluated numerically using an MCMC scheme to sample from the joint posterior distribution and make inference.  
\section{Derivation of Prior Distributions for Direct Sampling and Computational Improvements}
\label{comp}
Sampling from the joint posterior using MCMC methods is a tractable approach, but with come potential pitfalls, including poor mixing and identifiability issues particularly with the parameters $\delta_0$ and $\delta_1$.  In Bayesian methods it is preferable to first have a closed form for the posterior, allowing explicit analysis and inference, or second, to be able to sample directly from the joint posterior.  Existing prior distributions for spatial effects do not allow either of these approaches and instead rely on costly MCMC methods for evaluation.  This section presents a set prior distribution for spatial effects that allow direct sampling from the joint posterior.
\subsection{Derivation of the Posterior Distributions and Direct Sampling Sampling Scheme}
Given the likelihood \eqref{likelihood} the prior distributions \eqref{omega}, \eqref{fc0}, and \eqref{fc1} can be re-parametrised in by using the definition of $\eta=\delta_0/\delta_1$ and making the substitution $1/\delta_1=\eta/\delta_0$ into \eqref{omega}, the resulting prior distribution for $\bm{\nu}$ is   
\begin{eqnarray}
\label{eq:nuprior}
\pi\left(\bm{\nu}|\eta,\delta_0\right)&=&
|\bm{M}|^{1/2}_+\left(\frac{\eta}{2\pi\delta_0}\right)^{\frac{n-3}{2}}.
\end{eqnarray}
The prior distribution of $\delta_0$ remains as given in \eqref{delta}, and an arbitrary prior distribution $\pi(\eta)$ can be chosen subject to the constraint that $\eta>0$. 

The posterior of $\bm{\nu},\delta_0,\eta$ is
\begin{eqnarray}
\pi\left(\bm{\nu},\delta_0,\eta|\bm{y}\right)&\propto&
f(\bm{y}|\bm{\nu},\delta_0)\pi\left(\bm{\nu}|\eta,\delta_0\right)\pi\left(\delta_0\right)
\pi\left(\eta\right)
\end{eqnarray}
which has no closed form, but given the posterior distributions
\begin{eqnarray}
\label{post1}
\pi\left(\bm{\nu}|\delta_0,\eta,\bm{y}\right)\\
\label{post2}
\pi\left(\delta_0|\eta,\bm{y}\right)\\
\label{post3}
\pi\left(\eta|\bm{y}\right)
\end{eqnarray}
independent samples can be drawn from the joint posterior directly by exploiting the definition of the joint posterior distribution as
\begin{eqnarray}
\pi\left(\bm{\nu},\delta_0,\eta|\bm{y}\right)&=&
\pi\left(\bm{\nu}|\delta_0,\eta,\bm{y}\right)
\pi\left(\delta_0|\eta,\bm{y}\right)
\pi\left(\eta|\bm{y}\right).
\end{eqnarray}
The conditional posterior of $\bm{\nu}$ is given in \eqref{fcnu}, can be written in more compact notation as
\begin{eqnarray}
\label{eq:compactpost}
\bm{\nu}|\delta_0,\eta,\bm{y}&\sim&
N\left(\bm{A},\delta_0\bm{B}\right),\\
\bm{A}&=&\bm{B}\bm{y}\\
\bm{B}&=&\left(\bm{I}+
\eta\bm{M}\right)^{-1}
\end{eqnarray}
but the conditional posteriors 
\begin{eqnarray}
\label{int1}
\pi\left(\delta_0,\eta|\bm{y}\right)&\propto&
\int_{\mathbb{R}^n}f(\bm{y}|\bm{\nu},\delta_0)\pi\left(\bm{\nu}|\eta,\delta_0\right)\pi\left(\delta_0\right)
\pi\left(\eta\right)d\bm{\nu}\\
\label{int2}
\pi\left(\eta|\bm{y}\right)&\propto&
\int_{\mathbb{R}^+}\int_{\mathbb{R}^n}f(\bm{y}|\bm{\nu},\delta_0)\pi\left(\bm{\nu}|\eta,\delta_0\right)\pi\left(\delta_0\right)
\pi\left(\eta\right)d\bm{\nu}d\delta_0
\end{eqnarray}
are needed in order to complete the direct sampling scheme. 
Because \eqref{fcnu} is a proper density function  
\begin{eqnarray}
\label{eq1}
\int_{\mathbb{R}^n}\pi\left(\bm{\nu}|\delta_0,\eta,\bm{y}\right)d\bm{\nu}&=&1
\end{eqnarray}
then for \eqref{eq:compactpost}
\begin{eqnarray}
\label{eq2}
\int_{\mathbb{R}^n}\exp\left[-\frac{1}{2\delta_0}
\left(\bm{\nu}-\bm{A}\right)^T\bm{B}^{-1}
\left(\bm{\nu}-\bm{A}\right)\right]d\bm{\nu}&=&
\frac{(2\pi\delta_0)^{n/2}}{|\bm{B}|^{1/2}}.
\end{eqnarray}
Using this result and the fact that $\pi(\bm{\nu}|\delta_0,\eta,\bm{y})\propto f(\bm{y}|\delta_0,\bm{\nu})\pi(\bm{\nu}|\delta_0,\eta)$, written as:
\begin{eqnarray}
\label{eq3}
\pi\left(\bm{\nu}|\delta_0,\eta,
\bm{y}\right)&\propto&\exp\left\{-\frac{1}{2\delta_0}\left[\left(\bm{y}-\bm{\nu}\right)^T
\left(\bm{y}-\bm{\nu}\right)+\bm{\nu}^T\bm{M}\bm{\nu}
\right]\right\}.
\end{eqnarray}
the quadratic terms in \eqref{eq3} can be expanded the definitions of $\bm{A}$ and $\bm{B}$ substituted, yielding 
\begin{eqnarray}
\label{eq4}
\left(\bm{\nu}-\bm{A}\right)^T\bm{B}^{-1}
\left(\bm{\nu}-\bm{A}\right)&=&
\bm{\nu}^T\bm{B}^{-1}\bm{\nu}
-2\bm{\nu}^T\bm{B}^{-1}\bm{A}+
\bm{A}^T\bm{B}^{-1}\bm{A}
\end{eqnarray}
\begin{eqnarray}
\label{eq5}
\left[\left(\bm{y}-\bm{\nu}\right)^T
\left(\bm{y}-\bm{\nu}\right)+\bm{\nu}^T\bm{M}\bm{\nu}
\right]&=&\bm{y}^T\bm{y}-2\bm{\nu}^T\bm{y}+
\bm{\nu}^T\bm{\nu}+
\bm{\nu}^T\bm{M}\bm{\nu}
\end{eqnarray}
the term $\bm{A}^T\bm{B}^{-1}\bm{A}$ needs to be added to \eqref{eq5} to complete the square, and the term $\bm{y}^T\bm{y}$ can be factored out of the integral in \eqref{int1} yielding the solution for the posterior distribution \eqref{post2}.  Including the prior for $\delta_0$ \eqref{delta}, the the conditional posterior for $\delta_0,\eta$ is 
\begin{eqnarray}
\pi\left(\delta_0,\eta|\bm{y}\right)&\propto&
\frac{1}{\delta_0}
\left(\frac{\eta}{2\pi\delta_0}\right)^{\frac{n-3}{2}}
\frac{|\bm{M}|^{1/2}_+}
{|(\bm{I}+\eta\bm{M})|^{1/2}}
\exp\left[-\frac{1}{\delta_0}\left(\bm{y}^T\bm{y}-
\bm{y}^T(\bm{I}+\eta\bm{M})^{-1}\bm{y}\right)/2
\right].
\end{eqnarray}
resulting in the posterior for $\delta_0|\eta,\bm{y}$
\begin{eqnarray}
\label{postdelta}
\delta_0|\eta,\bm{y}&\sim&\mbox{InvGamma}
\left(a_0+\frac{n-3}{2},b_0+\left(\bm{y}^T\bm{y}-
\bm{y}^T(\bm{I}+\eta\bm{M})^{-1}\bm{y}\right)/2
\right).
\end{eqnarray}
From \eqref{int2}, the posterior distribution \eqref{post3} is 
\begin{eqnarray}
\pi(\eta|\bm{y})&\propto&
\pi(\eta)\int_{\mathbb{R}^+}\!
\frac{1}{\delta_0}
\left(\frac{\eta}{2\pi\delta_0}\right)^{\frac{n-t}{2}}
\!\!\!\!\!\!\frac{|\bm{M}|^{1/2}_+}
{|(\bm{I}+\eta\bm{M})|^{1/2}}
\exp\left[-\frac{1}{\delta_0}\left(\bm{y}^T\bm{y}-
\bm{y}^T(\bm{I}+\eta\bm{M})^{-1}\bm{y}\right)
\right]d\delta_0\\
\label{postlambda}
&\propto&
\pi(\eta)\times
\eta^{\frac{n-3}{2}}
\frac{|\bm{M}|^{1/2}_+}
{|(\bm{I}+\eta\bm{M})|^{1/2}}\frac{\Gamma\left(\frac{n-3}{2}\right)}
{\left[\left(\bm{y}^T\bm{y}-
\bm{y}^T(\bm{I}+
\eta\bm{M})^{-1}\bm{y}\right)/2\right]^{\frac{n-3}{2}}}
\end{eqnarray}
Samples can be drawn from \eqref{postlambda} using the ratio of uniforms method \citep{Kinderman:Monahan:1977}.  Then samples of $\delta_0$ and $\bm{\nu}$ can be drawn directly by substitution of samples of $\eta$ and $\delta_0$, resulting in a set of independent samples from the posterior distribution $\pi(\bm{\nu},\delta_0,\eta|\bm{y})$.    

The resulting joint posterior $\pi(\bm{\nu},\delta_0,\eta|\bm{y})$ can be sampled directly as follows:
\begin{enumerate}
\item Draw $n$ samples $\eta_1,\ldots,\eta_n$ from $\pi(\eta|\bm{y})$ 
\item Draw $n$ samples $\delta_{0,1},\ldots,\delta_{0,n}$ from  $\pi(\delta_{0,i}|\eta_i,\bm{y})$ 
\item Draw $n$ samples $\bm{\nu}_1,\ldots,\bm{\nu}_n$ from $\pi(\bm{\nu}_i|\delta_{0,i},\eta_i,\bm{y})$
\end{enumerate}
The resulting scheme doesn't require any ``burn-in'' and produces samples that are independent, making it more efficient than MCMC methods.  

\subsection{Computational Improvements for Sampling from  $\pi(\eta|\bm{y})$}

Efficient algorithms for drawing samples from the conditional distributions \eqref{fcnu} and \eqref{postdelta} are readily available, but drawing samples of $\eta|\bm{y}$ requires a bespoke solution, whose efficiency is dependent on reducing the computational burden of evaluating \eqref{postlambda}.  Initial inspection of \eqref{postlambda} reveals that there are several quantities 
$|\bm{M}|_+^{1/2}$, $\Gamma\left(\frac{n-3}{2}\right)$, and $\bm{y}^T\bm{y}$,
that only need to be computed once, thus can be ``pre-computed'' and stored in computer memory.  The remainder of the computational burden is in evaluating $\pi(\eta)$, which is assumed to be minimal, and evaluating the terms $|\bm{I}-\eta\bm{M}|^{1/2}$ and $\bm{y}^T(\bm{I}-\eta\bm{M})^{-1}\bm{y}$.  These quantities can be addressed by noting that the matrix $n$ by $n$ matrix $\bm{M}$ can be written
\[
\bm{M} = \bm{Q\Lambda Q}^T
\]
where $\bm{Q}$ is an $n$ by $n$ matrix of eigenvectors and $\bm{\Lambda}$ is a diagonal matrix of the eigenvalues $\lambda_1,\cdots,\lambda_n$. It can be shown that the eigenvalues of $\bm{I}+\eta\bm{M}$ are $1+\eta\lambda_1,\cdots,1+\eta\lambda_n$, and that
\[
|\bm{I}+\eta\bm{M}|^{1/2}=\left(\prod_{i=1}^n(1+\eta\lambda_i)\right)^{1/2}.
\]
The identity $\bm{QQ}^T=\bm{I}$ is used to write 
\begin{eqnarray}
\bm{y}^T(\bm{I}-\eta\bm{M})^{-1}\bm{y}&=&
\bm{y}^T(\bm{I}+\eta\bm{Q\Lambda Q}^T)^{-1}\bm{y}\\[9pt]
&=&\bm{y}^{T}\bm{Q}
\left(
\begin{array}{ccc}
\frac{1}{1+\eta\lambda_1} &  \cdots & 0\\
0 & \ddots  & 0 \\
0 & \cdots  & \frac{1}{1+\eta\lambda_n}
\end{array}\right)\bm{Q}^T\bm{y}.
\end{eqnarray}
The eigenvalues, eigenvectors, and the vector $\bm{y}^*=\bm{Q}^T\bm{y}$ can all be ``pre''-computed, offering a significant reduction in computational cost for evaluating $\pi(\eta|\bm{y})$ or $\log(\pi(\eta|\bm{y}))$, depending on the requirements of the chosen sampling method.  A similar approach to this is demonstrated in \citep{He:Sun:2000} and \citep{White:2006}.   
\section{Numerical Examples}
\label{examples}
The efficiencies resulting from the direct sampling scheme derived in Section 3 are illustrated using sample datasets.   In the first example, samples are drawn directly from the joint posterior distribution, and are independent resulting in effective sample sizes equal to the number of draws.  In the second example, the direct sampling scheme is implemented in a Markov Chain Monte Carlo scheme to evaluate results from a hierarchical model with a non-Gaussian likelihood.  In this case the ability to draw joint samples from a subset of the parameters improves sampling efficiency substantially, reducing the total number of iterations needed to obtain a desired effective sample size.  

\subsection{Meuse River Data}
The Meuse river data set from \citep{Burrough:McDonnell:1998}, included in the \texttt{sp} package \citep{Pebesma:Bivand:2005,Bivand:etal:2013}  for R \citep{R}, contains measurements of heavy metal concentrations (cadmium, copper, lead, and zinc) in the topsoil of a flood plain at 155 locations along the Meuse river in France.  As assumed in the vignette for the \texttt{sp} package, the concentrations can be assumed to follow a log-normal distribution.  
\begin{eqnarray}
\log(\bm{y})&\sim&N(\bm{\nu},\delta_0\bm{I})
\end{eqnarray}
Defining $\bm{y}^*=\log(\bm{y})$ and based on (13) the resulting likelihood for the data is
\begin{eqnarray}
\bm{y}^*&\sim&N(\bm{\nu},\delta_0\bm{I}),
\end{eqnarray}
and the following priors complete the model
\begin{eqnarray}
\left[\bm{\nu}\mid\eta,\delta_0\right]&\propto& \frac{\eta}{\delta_0}^{-(n-3)/2}
\exp\left(-\frac{\eta}{2\delta_0}\bm{\nu}'\bm{M}\bm{\nu}\right)\\
\delta_0&\sim&\mbox{InvGamma}(a_0,b_0)\\
\pi(\eta)&=&\frac{1}{(1+\eta)^2}.
\end{eqnarray}
Note that the prior for $\eta$ is a Pareto density with an undefined mean and variance, and a median of 1. It is also the equivalent of defining $q$=$\eta/(1+\eta)$ and putting a uniform prior on $q$ over the interval $(0,1)$, i.e.\  $q\sim U(0,1)$.  The choice of prior for $\eta$ is somewhat computationally arbitrary as the burden of evaluating $\pi(\eta)$ is a small portion of the computational cost of evaluating \eqref{postlambda}.

\begin{figure}
\caption{\label{meuse:post}Posterior Densities of $\delta_0$ (left) and $\eta$ (right) }
\vspace{3mm}
\includegraphics[width=0.5\textwidth]{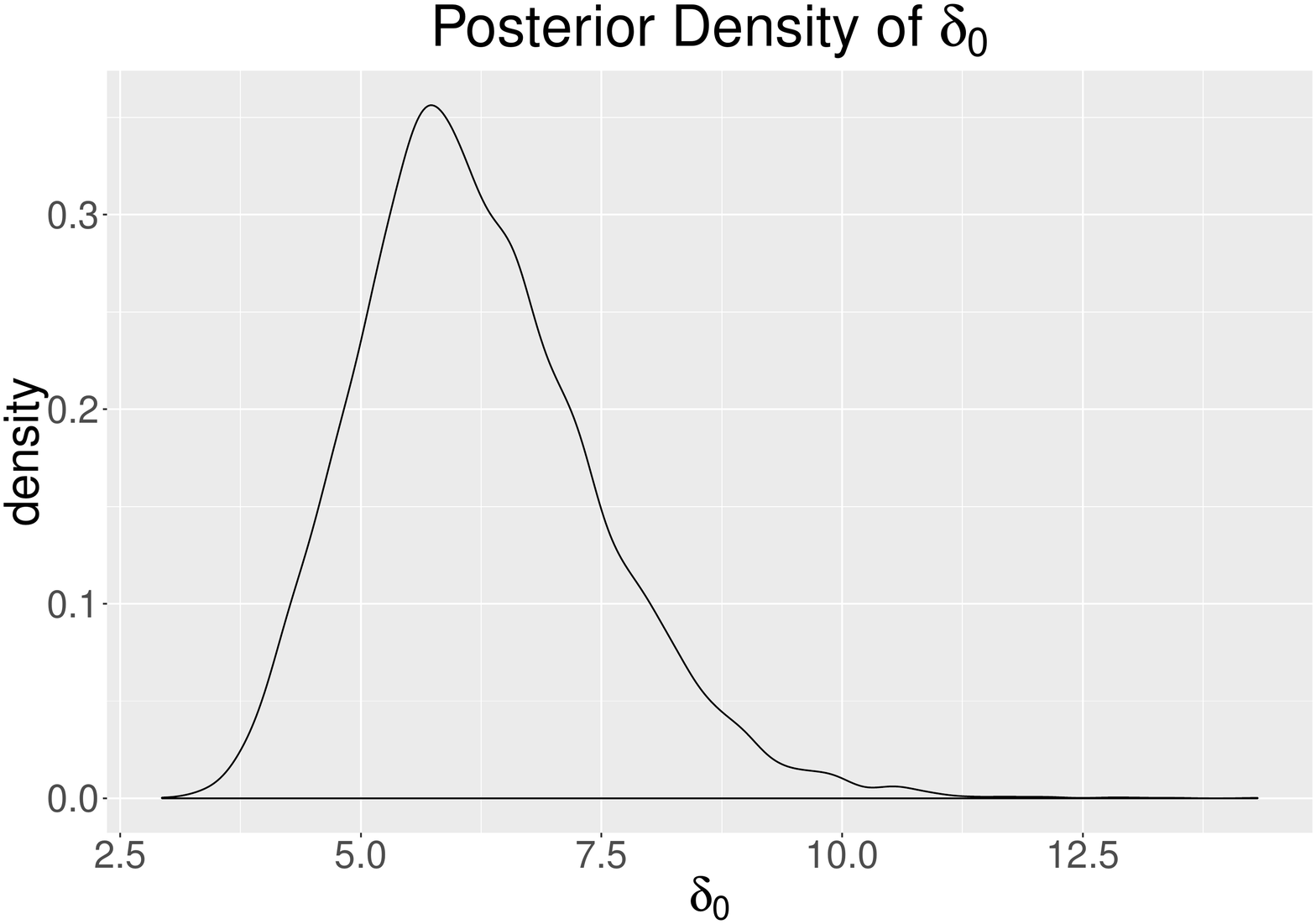}
\includegraphics[width=0.5\textwidth]{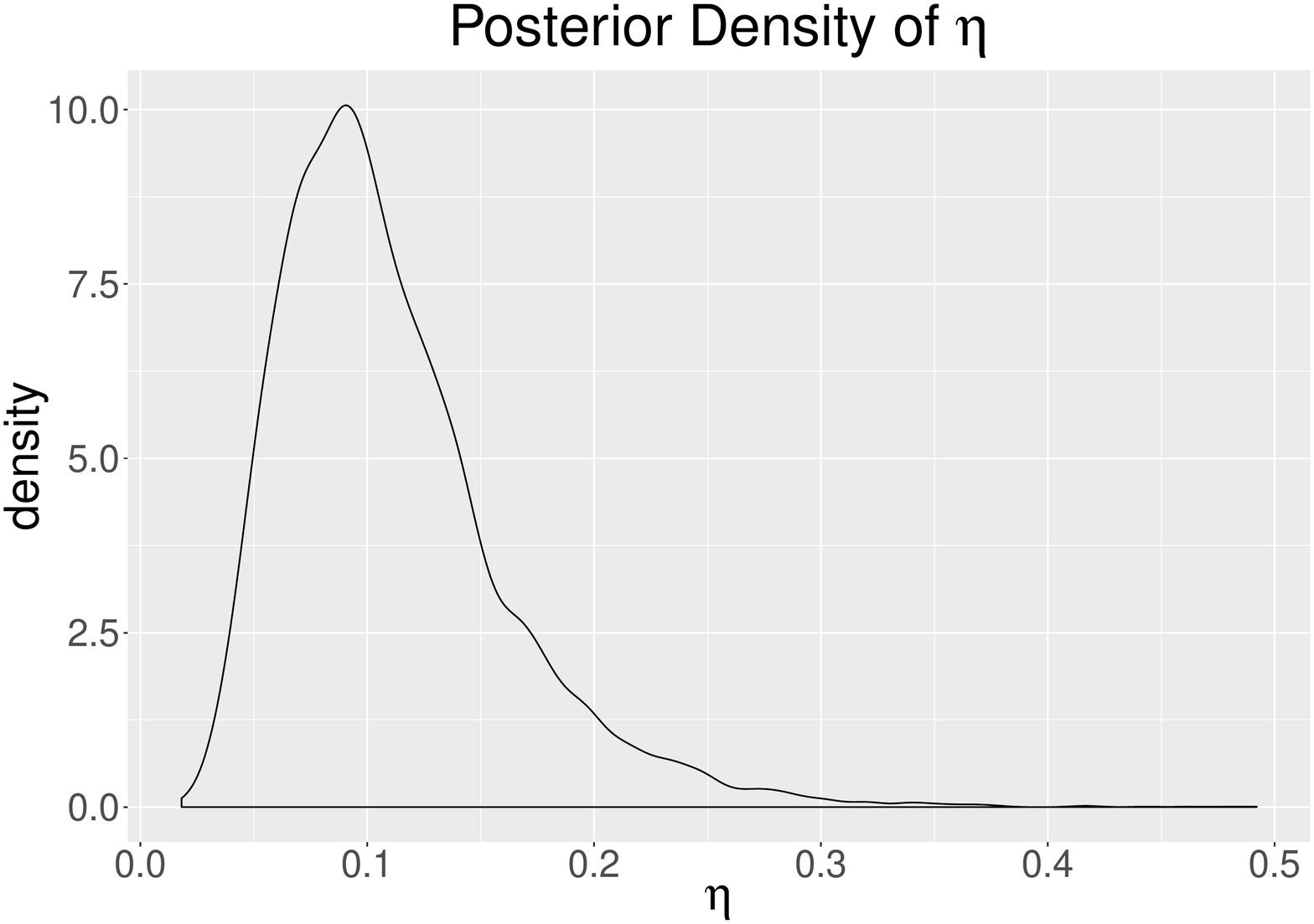}\\
\end{figure}

\begin{figure}
\caption{\label{meuse:acf}Auto-Correlation Functions for $\delta_0$ (left) and $\eta$ (right) }
\vspace{3mm}
\includegraphics[width=0.5\textwidth]{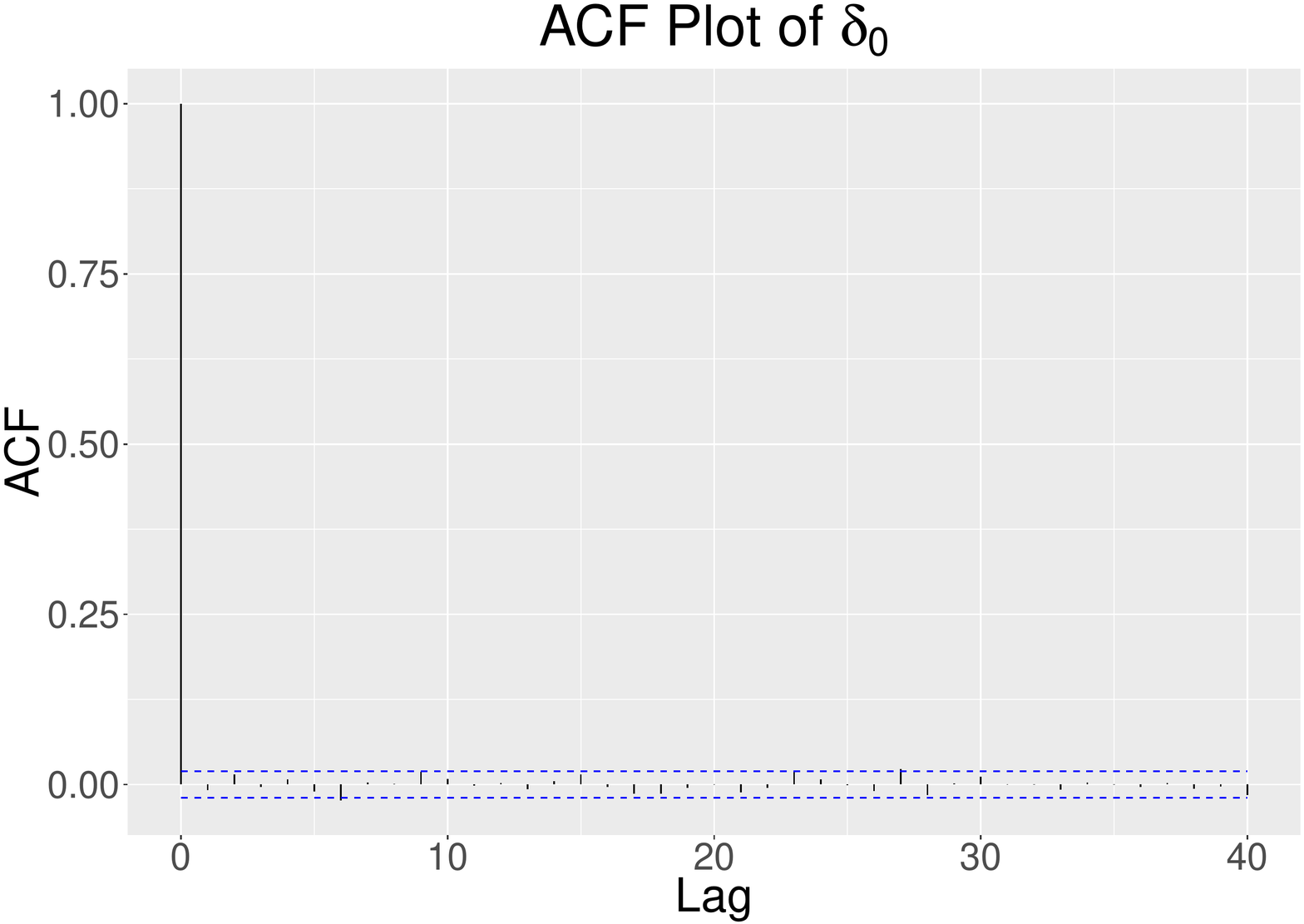}
\includegraphics[width=0.5\textwidth]{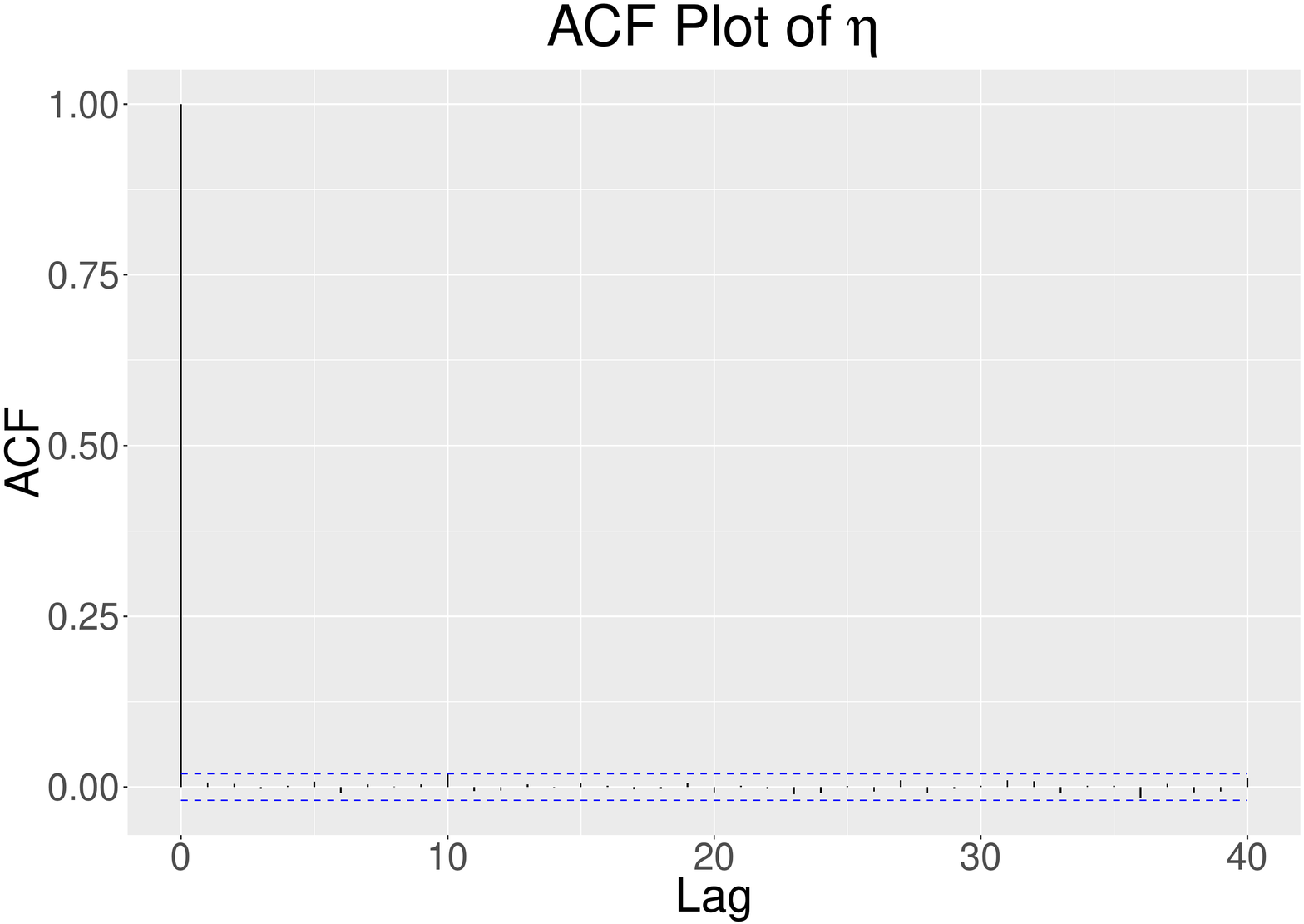}
\end{figure}

The resulting model can be evaluated as defined in Section \ref{comp}, drawing independent samples directly from the full posterior.  Running R 3.3.3 on a iMac mini with 16 GB of RAM and an 3.0 GHz Intel Core i7 processor 10,000 samples are drawn in 6.3 seconds.
The posterior densities for $\delta_0$ and $\eta$ in Figure \ref{meuse:post} appear smooth, and auto-correlation functions for $\delta_0$ and $\nu$ in Figure \ref{meuse:acf} show no evidence of correlation between draws.  

\begin{figure}
\caption{\label{meuse:data}Heat Maps of the Observed Data $y^*$ (left) and Smoothed values $\bm{\nu}$ (right)}
\vspace{3mm}
\includegraphics[width=0.5\textwidth]{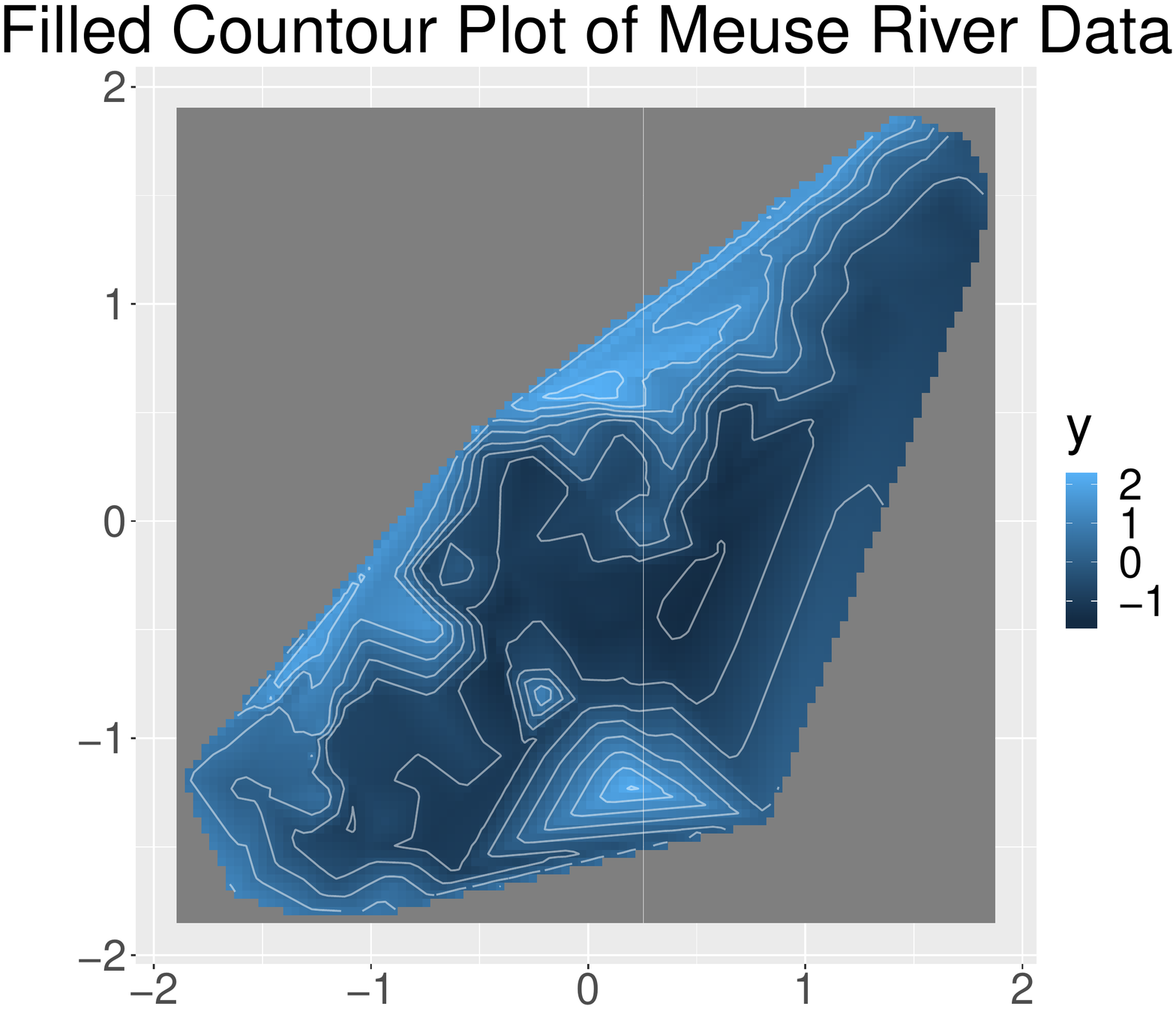}
\includegraphics[width=0.5\textwidth]{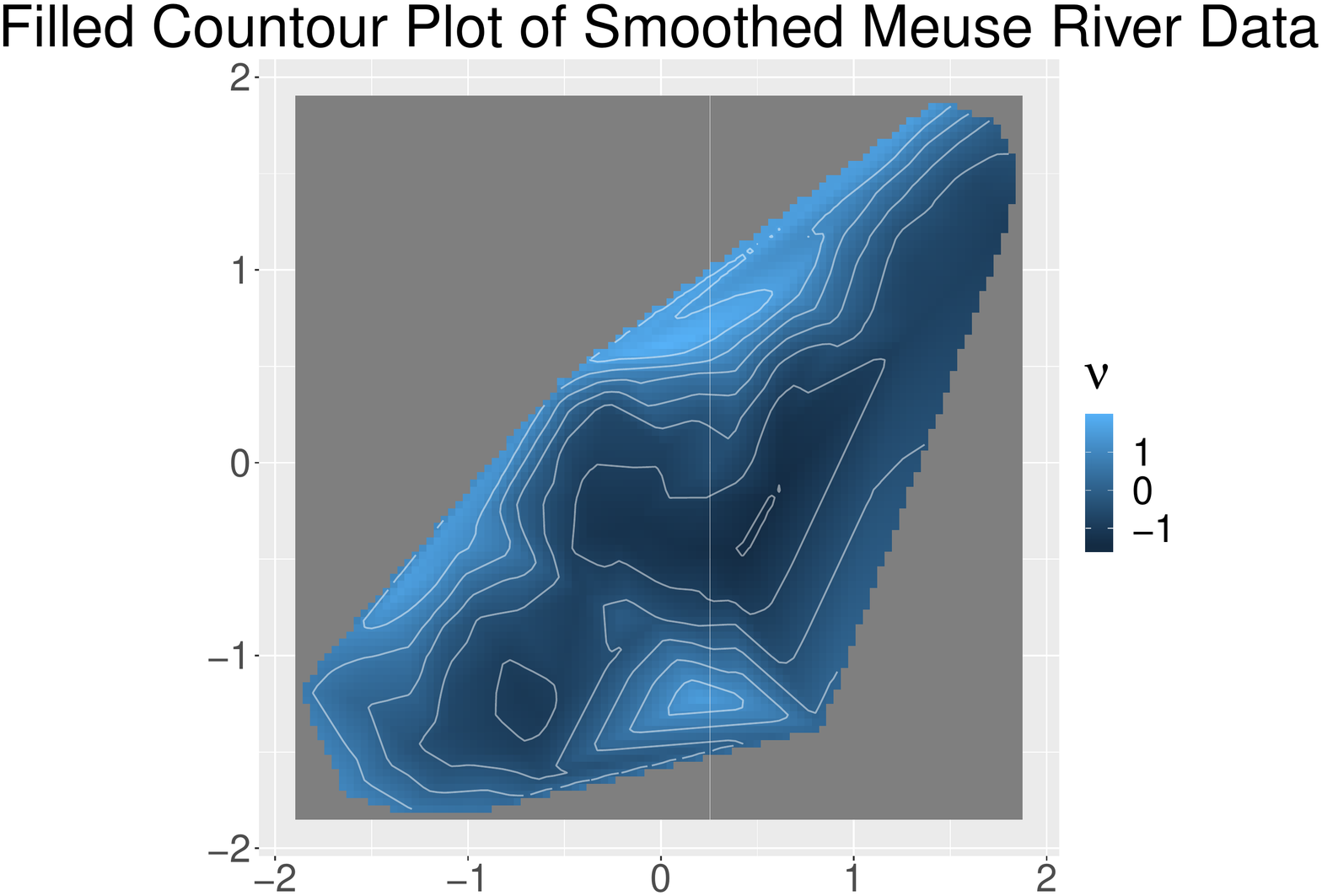}
\end{figure}

\subsection{Missouri Turkey Hunting Survey}
The results derived in Section \ref{comp} are straightforward, considering data from a Gaussian likelihood.  In practice spatial smoothing occurs in a wide variety of cases as part of a generalised linear mixed effects model (GLMM) or in the Bayesian interpretation a hierarchical model \citep{Banerjee:2016}.  One example of this type of model is suggested in \citep{He:Sun:2000} for data concerning hunters' success rates in the Missouri turkey hunting season of 1996.  

In 1996 the Missouri Department tested using postal surveys to elicit the information on where and when hunters hunted, and if they were successful. The resulting surveys collected information for most of the 114 counties in Missouri for both weeks of the hunting season.  The resulting data provide a useful dataset to illustrate the use of the direct-sampling spatial prior in a hierarchical model. 
If $y_{ij}$ is equal to the number of turkeys harvested in county $i=1,\ldots,N=114$ in week $j=1,2$, and $n_{ij}$ are the number of individuals who hunted in county $i$ during week $j$, then as per \citep{He:Sun:2000}
\begin{eqnarray}
\label{eq:mthslike}
y_{ij}\sim Binom(n_{ij},p_{ij}).
\end{eqnarray}
The hierarchical model is created by defining $p_{ij}$ as
\begin{eqnarray}
\label{eq:mthseq1}
\nu_{ij}\equiv\left(\frac{p_{ij}}{1-p_{ij}}\right)=
Z_i+\theta_j+\epsilon_{ij}
\end{eqnarray}
and 
\begin{eqnarray}
\label{eq:eps}
\epsilon\stackrel{iid}{\sim}N(0,\delta_0).
\end{eqnarray}
The term $Z_i$ represents the spatial effects and has a prior as in \eqref{eq:nuprior}, the term $\theta_j$ is the difference between weeks 1 and 2, hence $\theta_1=0$.  Assigning a flat prior for $\theta_2$ and following the derivation in \eqref{post1} --\eqref{postlambda} as set of posterior distributions can be derived 
\begin{eqnarray}
\pi(\bm{Z}|\delta_0,\eta,\theta_2,\bm{\nu})\\
\pi(\delta_0|\eta,\bm{\nu})\\
\pi(\eta|\bm{\nu})
\end{eqnarray}
Which will allow for direct sampling from the posterior of$\bm{Z}=(Z_1,\ldots,Z_{N})'$, given $\bm{\nu}$ and $\theta_2$.  

The complete model will still need to be evaluated in an MCMC scheme, because the conditional posterior of $\bm{\nu}$ is only known up to a proportionality constant with no closed form, 
\begin{eqnarray}
\pi(\nu_{ij}|Z_i,\theta_j,\delta_0,\bm{y})\propto
\exp\left\{\nu_{ij}y_{ij}-n_{ij}\log\left(1+e^{\nu_{ij}}\right)-
\frac{1}{2\delta_0}\left(\nu_{ij}-Z_i-\theta_j\right)^2\right\}.
\end{eqnarray}
 and $\theta_2$ can't be integrated out as part of the direct sampling scheme though it does have a closed form for its conditional posterior distribution
\begin{eqnarray}
(\theta_2|\bm{Z},\delta_0,\bm{\nu})\sim N\left(\frac{\sum_{i=1}^{N}(\nu_{i2}-Z_i)}{N},\frac{\delta_0}{N}\right).
\end{eqnarray}
By comparison,  the full-conditional posterior distributions
\begin{eqnarray}
\pi(\bm{\nu}|\delta_0,\theta_2,\bm{Z}),\bm{y})\\
\pi(\bm{Z}|\delta_0,\eta,\theta_2,\bm{\nu})\\
\pi(\theta_2|\delta_0,\bm{nu},\bm{Z})\\
\pi(\delta_0|\theta_2,\bm{Z},\bm{\nu})\\
\pi(\eta|\delta_0,\theta_2,\bm{Z})
\end{eqnarray}
cam be easily derived and used to construct a more traditional MCMC sampling scheme.  

\begin{figure}
\begin{center}
\caption{Auto-correlation Function Plots for Direct Sampling Scheme (left) and Traditional MCMC (right)\label{figure1}}
\vspace{3mm}
\includegraphics[width=0.45\textwidth]{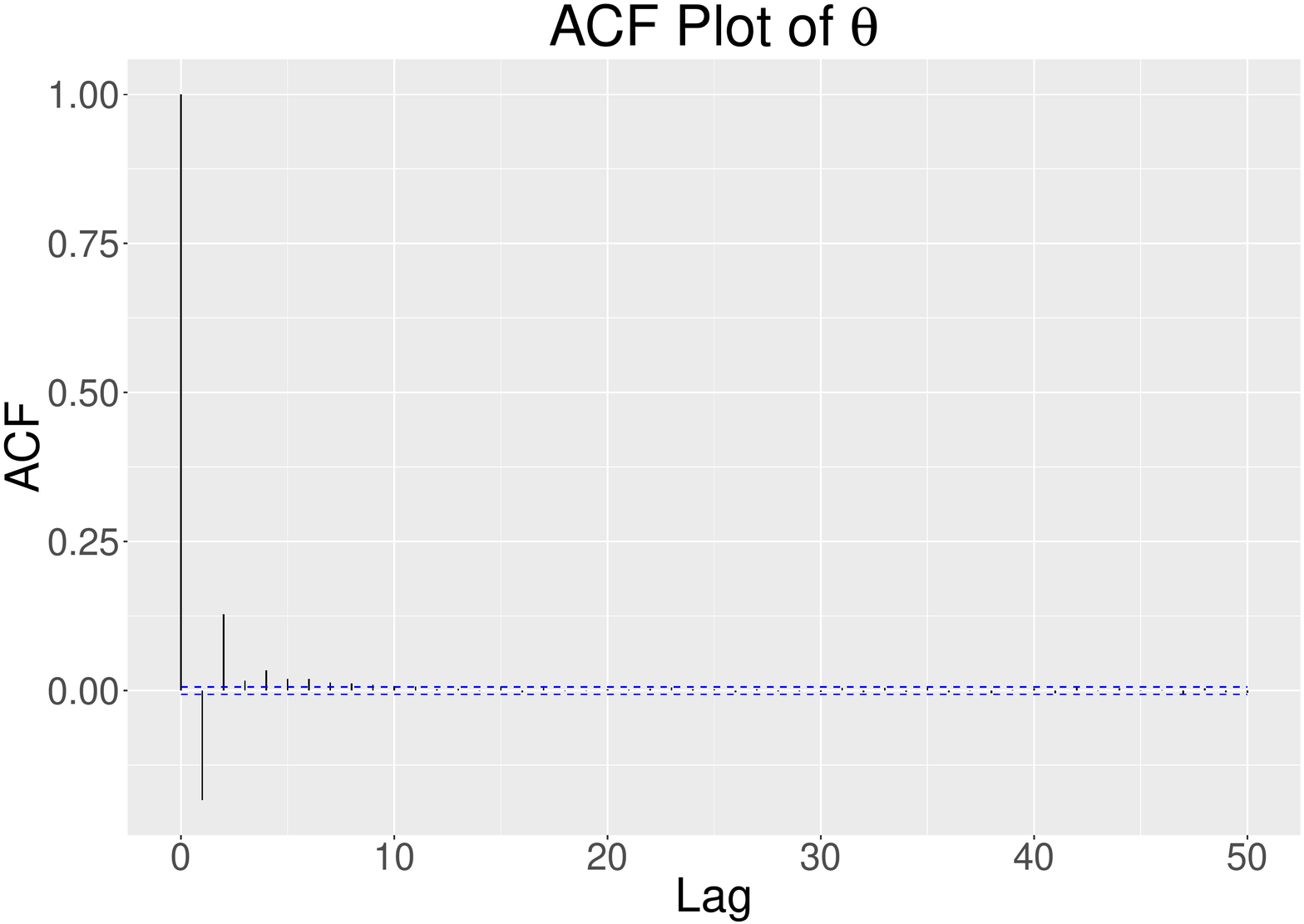}
\includegraphics[width=0.45\textwidth]{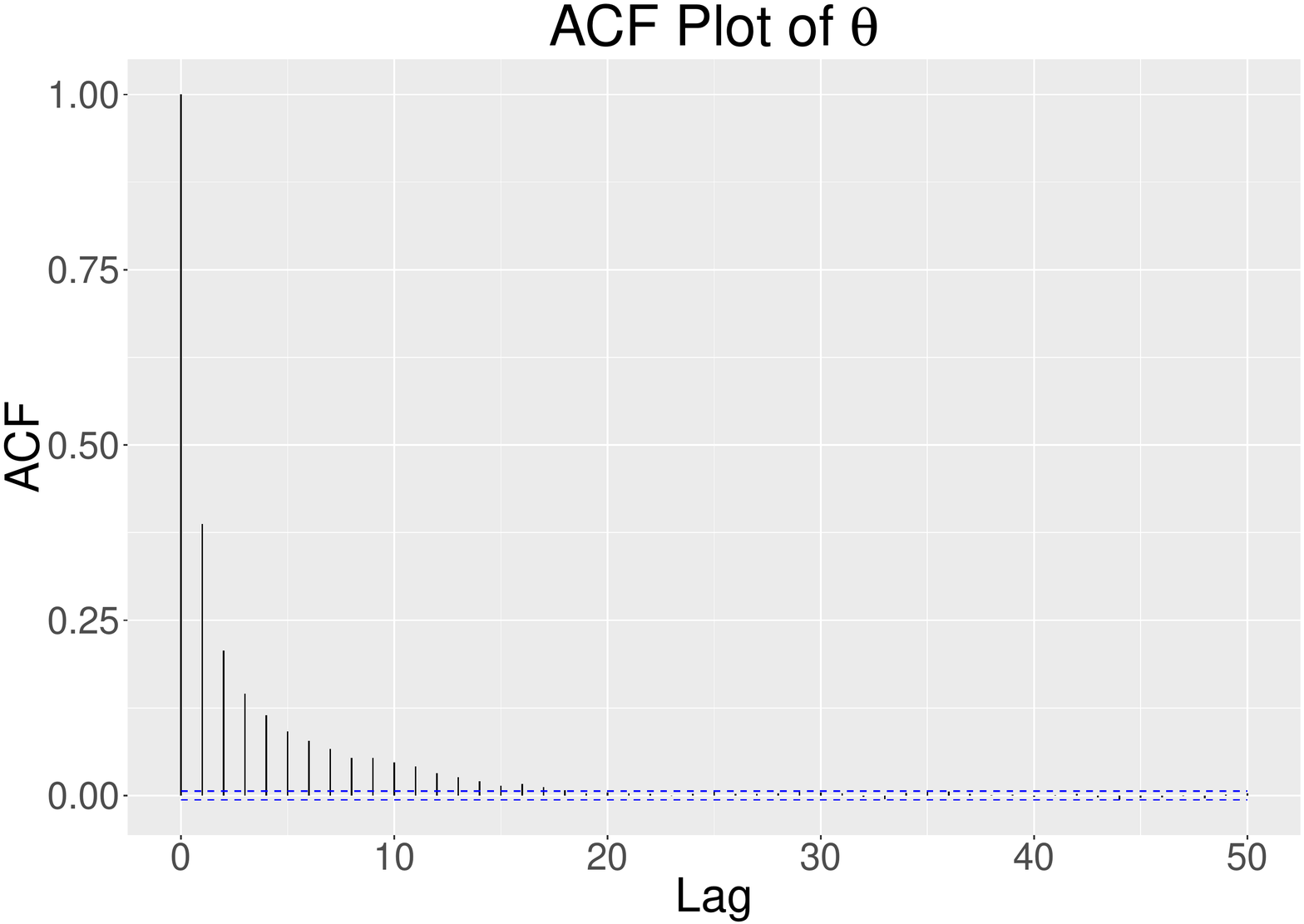}\\
\includegraphics[width=0.45\textwidth]{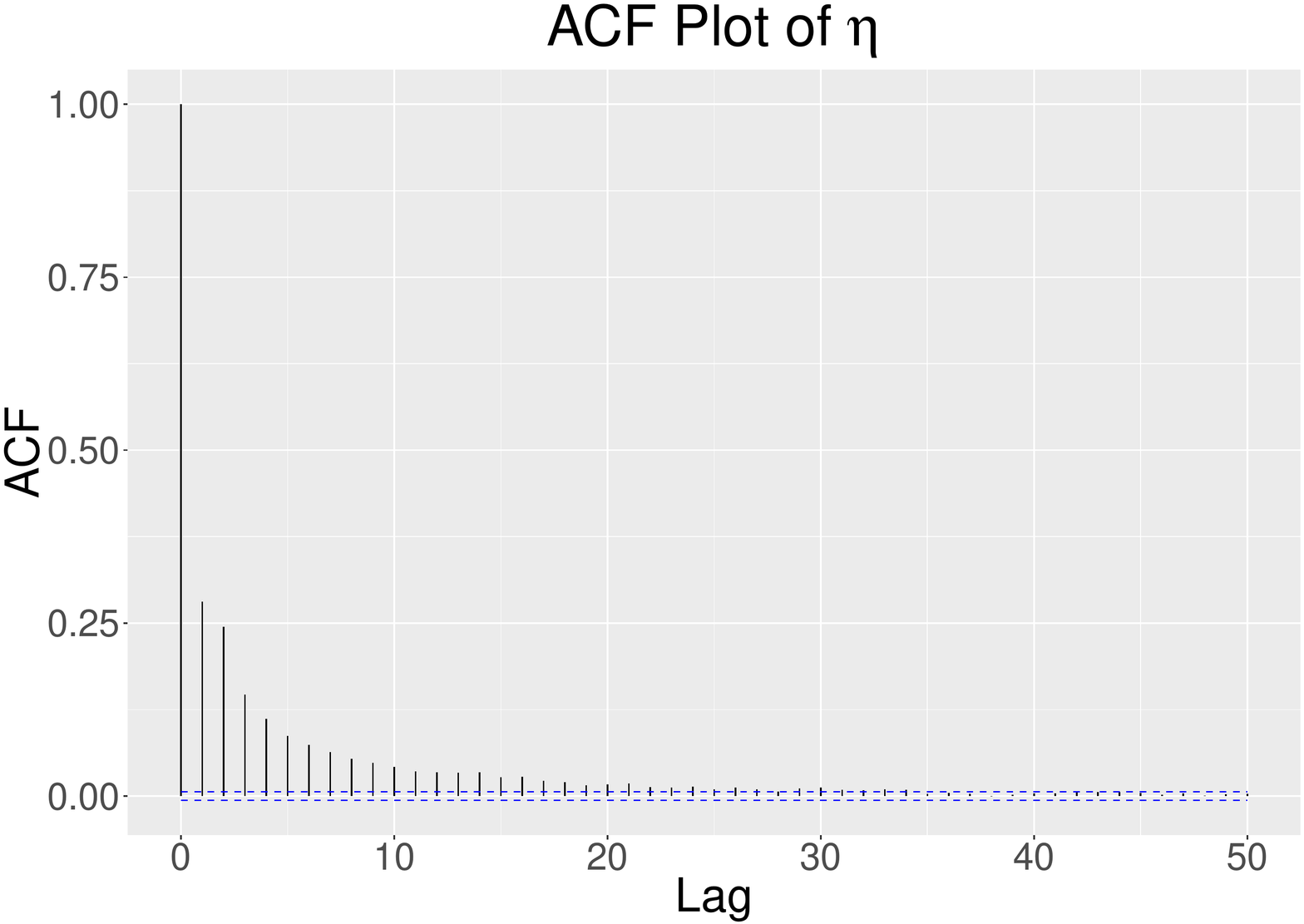}
\includegraphics[width=0.45\textwidth]{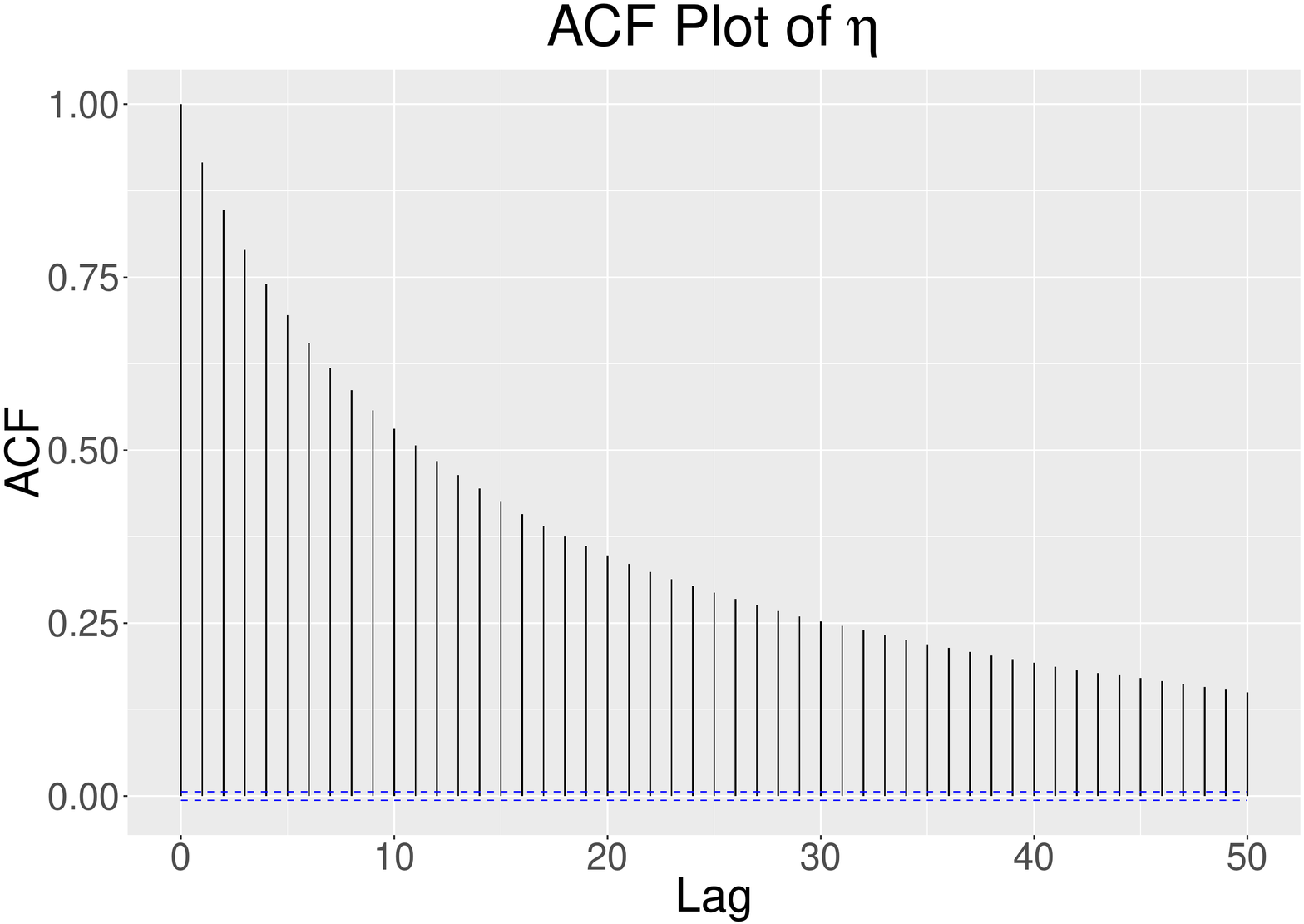}\\
\includegraphics[width=0.45\textwidth]{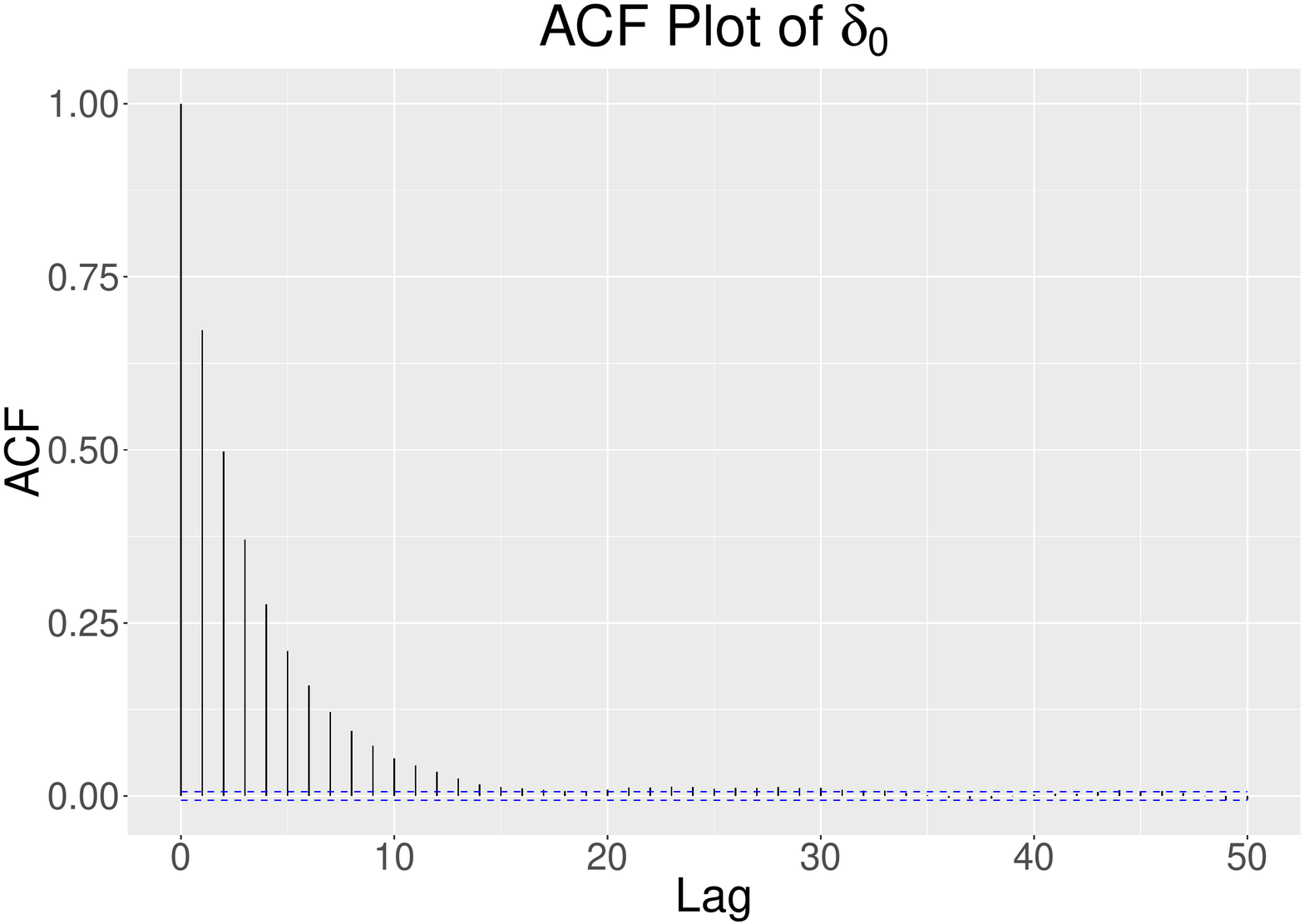}
\includegraphics[width=0.45\textwidth]{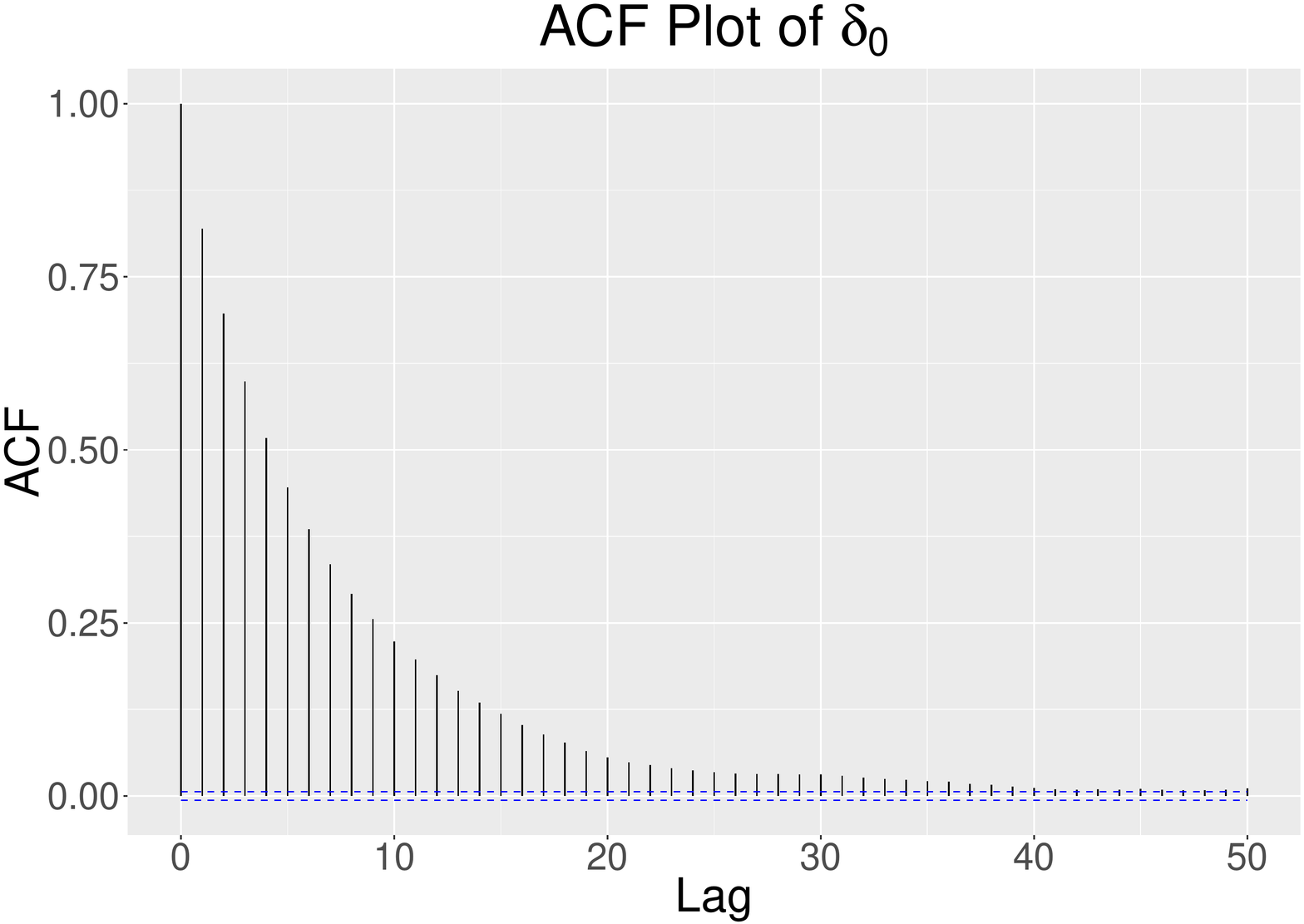}\\
\end{center}
\end{figure}

Using the direct sampling scheme for $\bm{Z},\delta_0,\eta$ should however provide an improvement in efficiency as measured by the effective number of samples resulting from draws from the full conditional posterior, as calculated in \citep{Gong:Flegal:2015}.  Despite the increased execution time the direct sampling scheme compared to the traditional MCMC scheme using the full conditionals (13 minutes, 20 seconds to 16 minutes, 5 seconds respectively) results in Table \ref{tab1} show that, as expected, the direct sampling scheme does result in a larger effective sample size.  The resulting plots of the auto-correlation factors in Figure \ref{figure1} provide further illustration of the increased efficiency of the direct sampling scheme.  

\begin{table}
\centering
\caption{Comparison of Effective Sample Sizes for Full Conditional and Direct Sampling Schemes from $100,\!000$ Iterations\label{tab1}}
\vspace{3mm}
\begin{tabular}{cccccc}
\toprule
\multicolumn{3}{c}{Traditional MCMC}& 
\multicolumn{3}{c}{Direct Sampling Spatial Prior}\\
\midrule
$\theta_2$ & $\eta$ & $\delta_0$ & $\theta_2$ &$\eta$ & $\delta_0$\\
\midrule
24,893.86& 2153.90 & 6840.54 & 81,148.28 & 24,844.01 & 15,544.92\\
\bottomrule
\end{tabular}
\end{table}

\begin{figure}
\begin{center}
\caption{Posterior Density Function Plots for Direct Sampling Scheme (left) and Traditional MCMC (right)\label{fig:density}}
\vspace{3mm}
\includegraphics[width=0.45\textwidth]{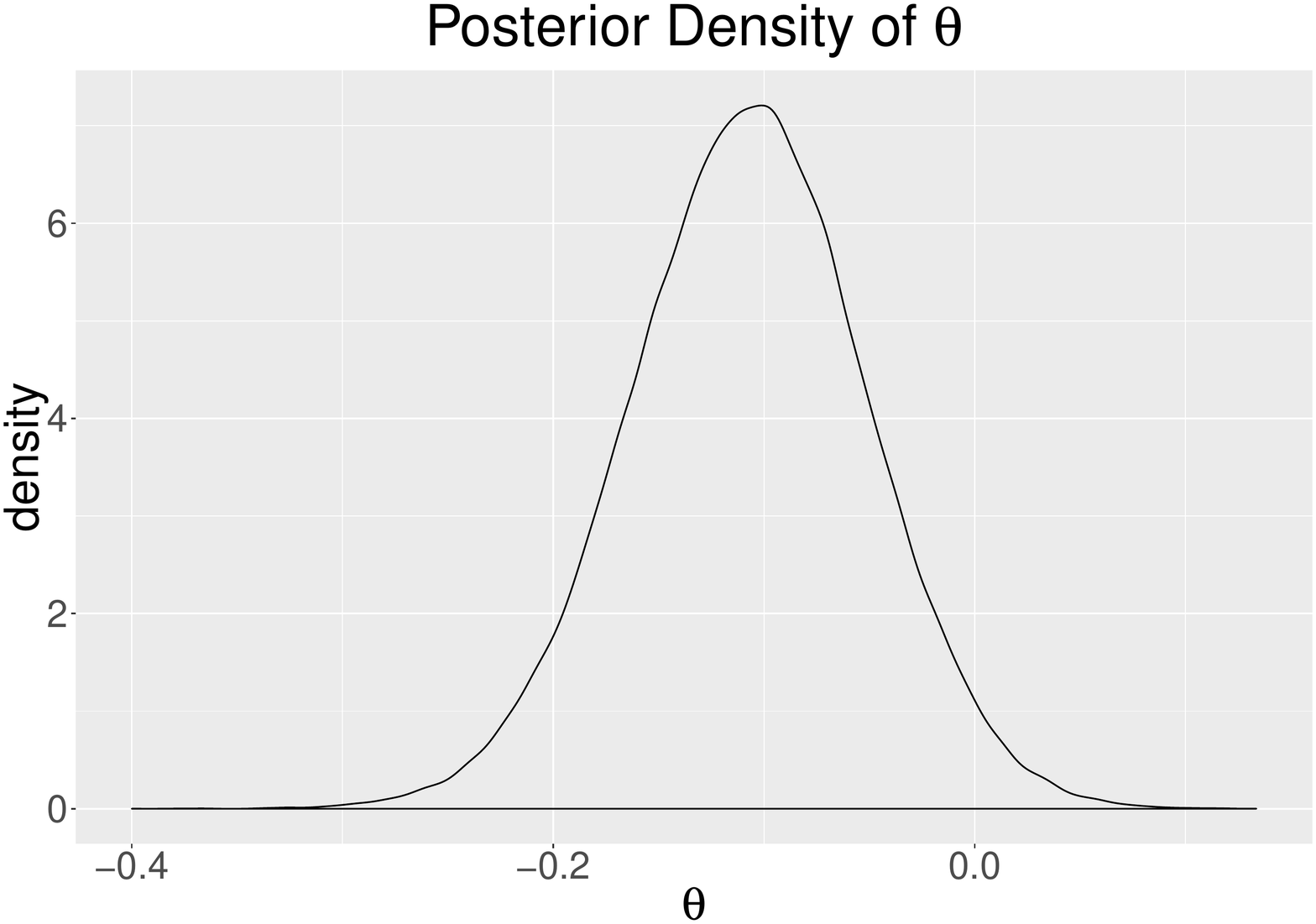}
\includegraphics[width=0.45\textwidth]{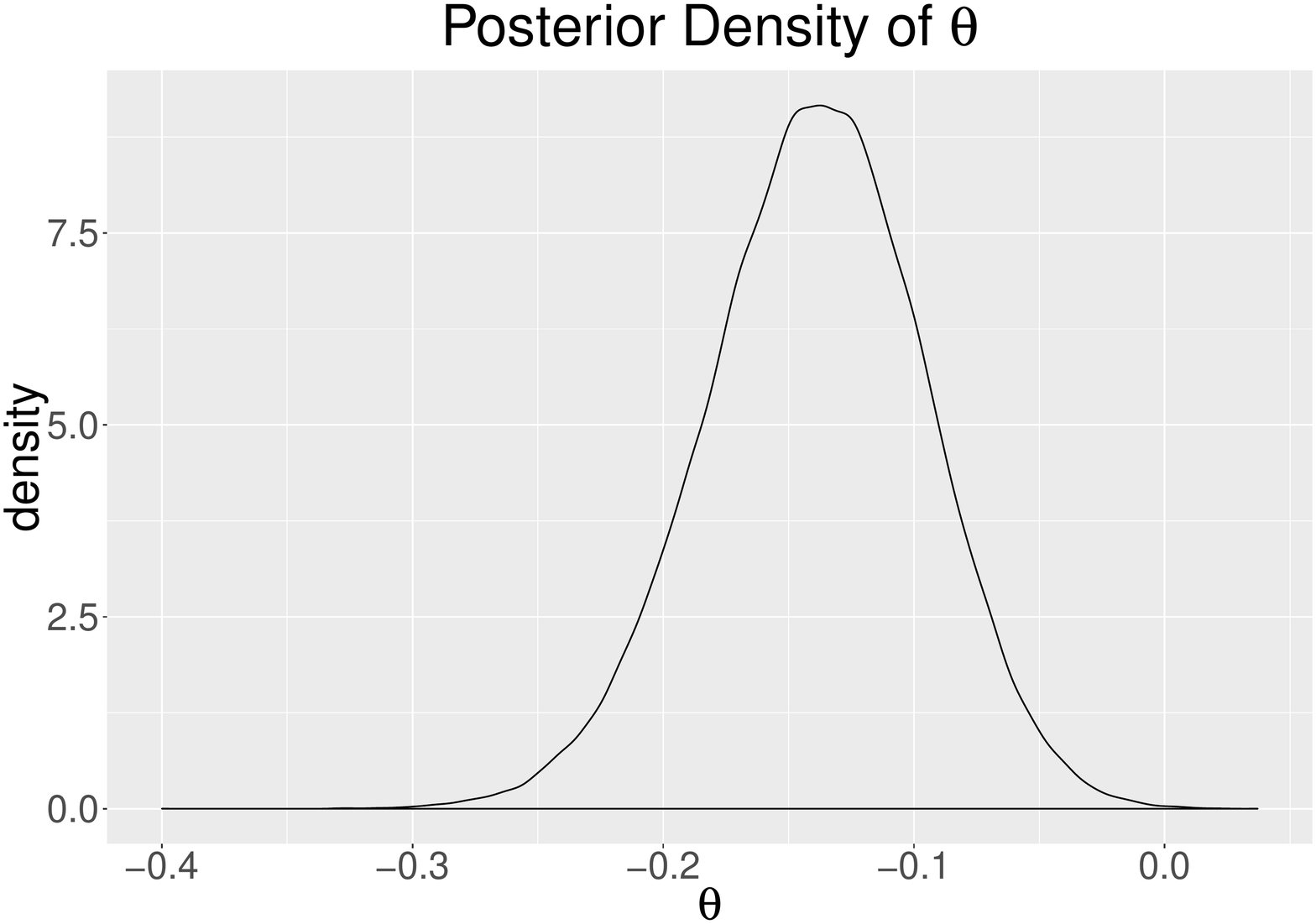}\\
\includegraphics[width=0.45\textwidth]{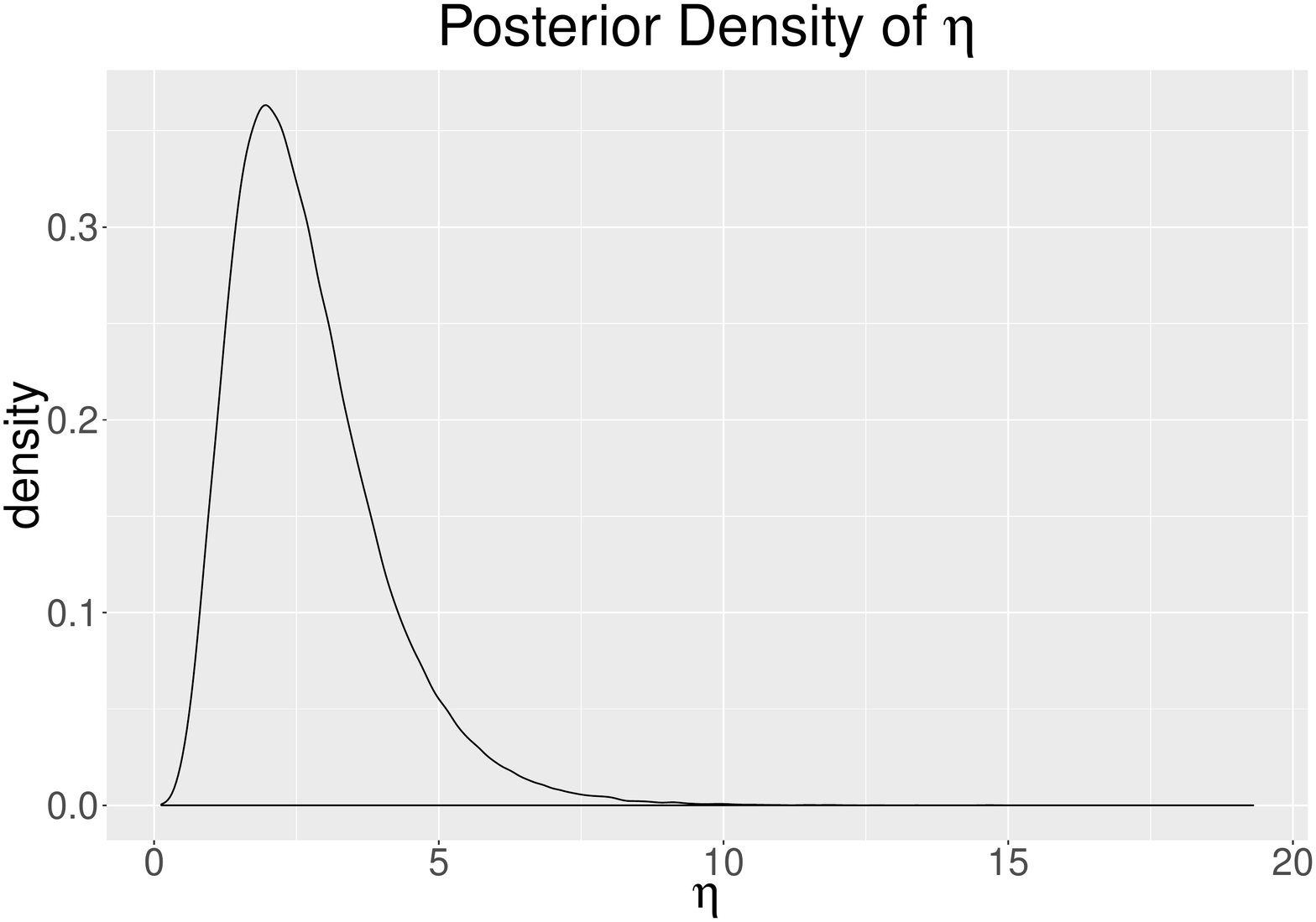}
\includegraphics[width=0.45\textwidth]{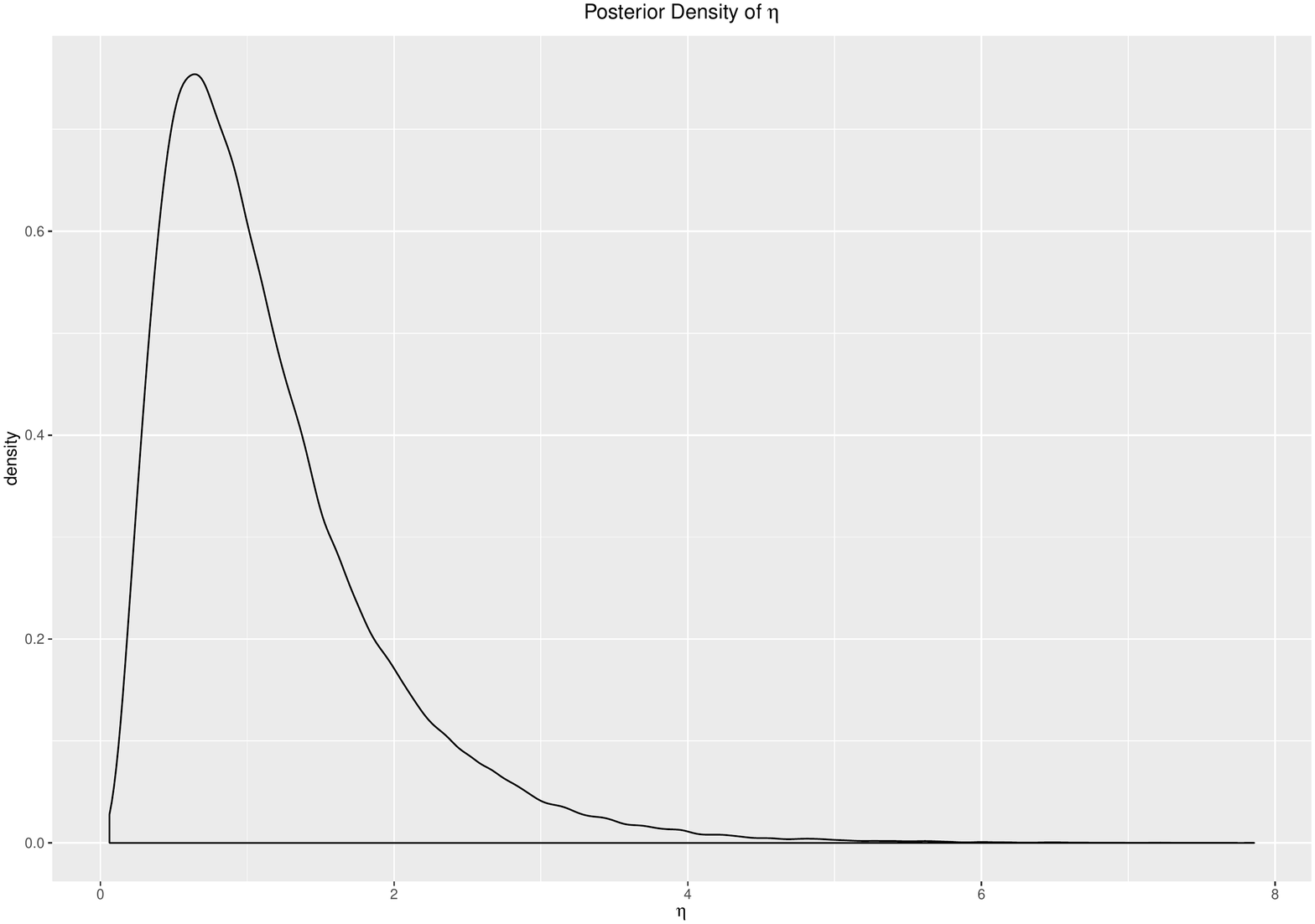}\\
\includegraphics[width=0.45\textwidth]{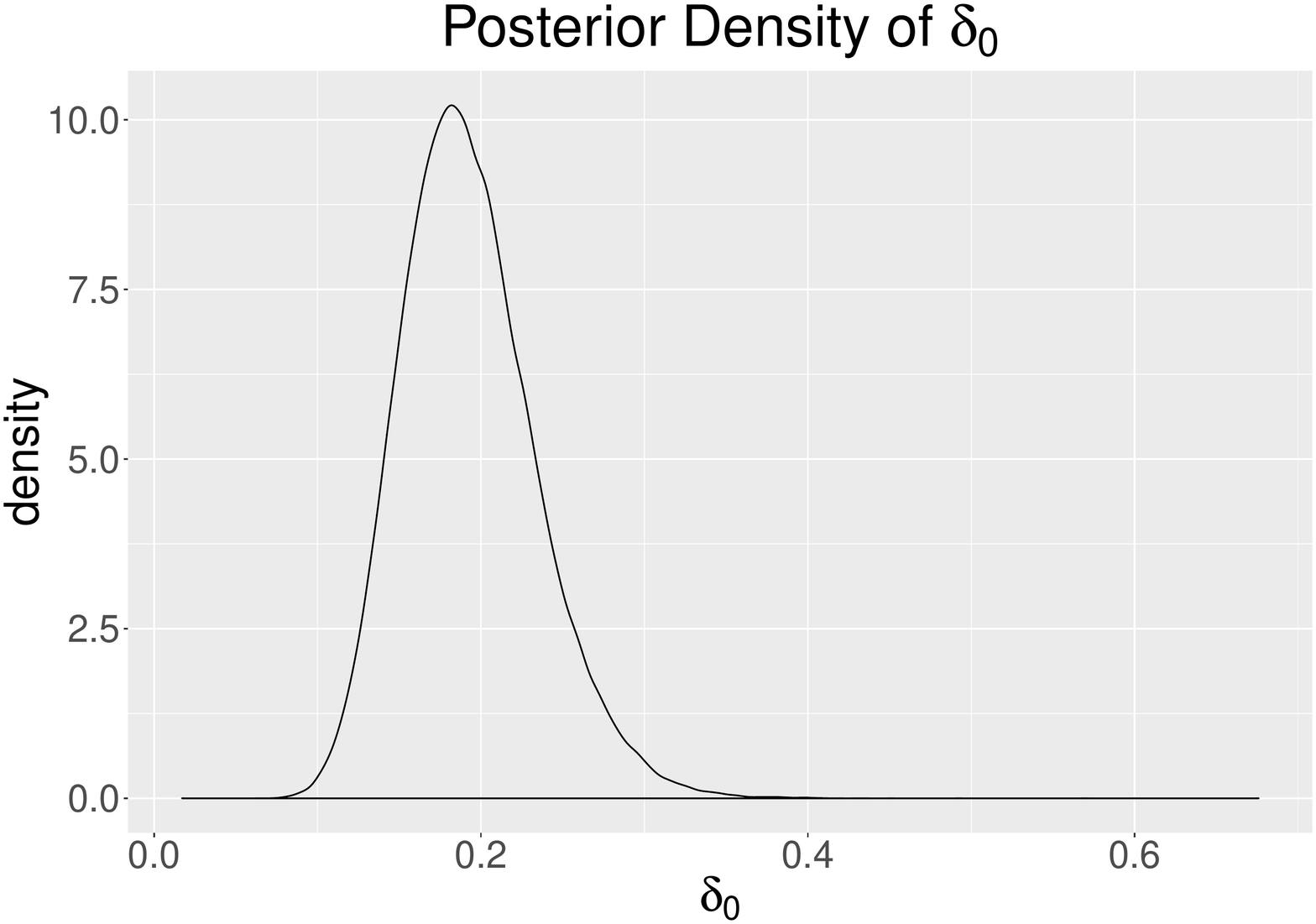}
\includegraphics[width=0.45\textwidth]{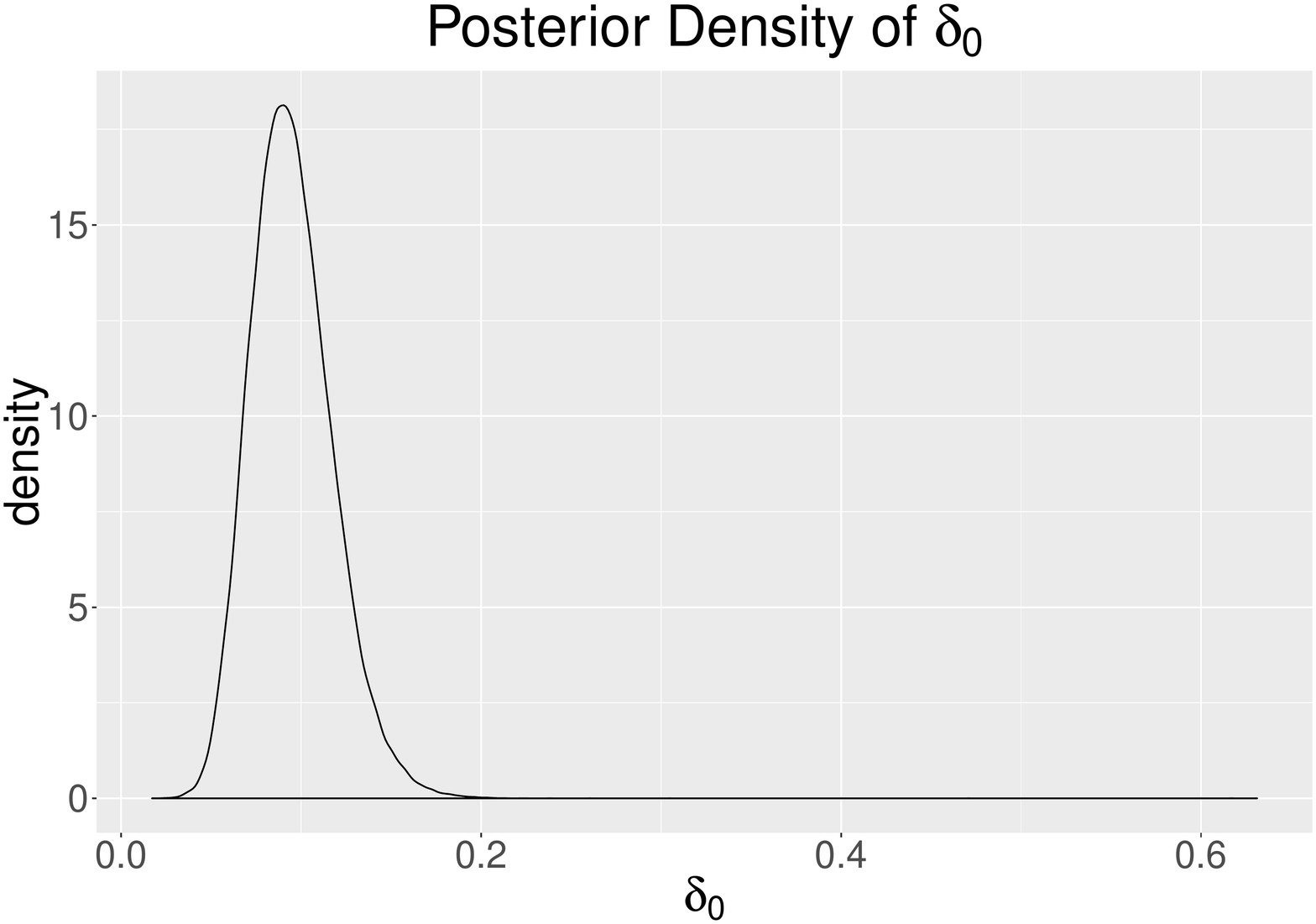}
\end{center}
\end{figure}
\section{Discussion}
\label{disc}
This paper presents a prior distribution for spatial random effects based on using thin-plate splines radial basis functions and boundary conditions. The resulting prior is improper but has a Gaussian form and yields a proper posterior distribution for spatial effects.  If the data follow a Gaussian likelihood then resulting posterior density of the model parameters (including spatial effects)  can be sampled directly without use of an MCMC scheme resulting in much shorter model evaluation times, nearing those of classical modelling approaches, or posterior approximation methods.  In the case of non-Gaussian data the direct sampling scheme can be extended to apply to hierarchical or generalised linear mixed effects models resulting in more efficient sampling resulting from reduced correlation in MCMC chains and increased effective sample sizes.  The computational benefit of this approach is not limited to spatial priors.  The use of thin-plate splines basis functions can be used to construct similar prior distributions over higher dimensioned space for use in a broad range of smoothing problems.


\begin{thebibliography}{34}
\providecommand{\natexlab}[1]{#1}
\providecommand{\url}[1]{\texttt{#1}}
\expandafter\ifx\csname urlstyle\endcsname\relax
  \providecommand{\doi}[1]{doi: #1}\else
  \providecommand{\doi}{doi: \begingroup \urlstyle{rm}\Url}\fi

\bibitem[Banerjee(2016)]{Banerjee:2016}
Sudipto Banerjee.
\newblock Spatial data analysis.
\newblock \emph{Annual Review of Publich Health}, 37:\penalty0 47--60, 2016.

\bibitem[Banerjee et~al.(2015)Banerjee, Carlin, and
  Gelfand]{Banerjee:etal:2015}
Sudipto Banerjee, Bradley~P. Carlin, and Alan~E. Gelfand.
\newblock \emph{Hierarchical Modeling and Analysis for Spatial Data}.
\newblock CRC Press, 2nd edition, 2015.

\bibitem[Besag(1974)]{Besag:1974}
Julian Besag.
\newblock {S}patial {I}nteraction and the {S}tatistical {A}nalysis of {L}attice
  {S}ystems.
\newblock \emph{Journal of the Royal Statistical Society, Series B},
  36:\penalty0 192--236, 1974.

\bibitem[Bivand et~al.(2013)Bivand, Pebesma, and Gomez-Rubio]{Bivand:etal:2013}
Roger~S. Bivand, Edzer Pebesma, and Virgilio Gomez-Rubio.
\newblock \emph{{A}pplied Spatial Data Analysis with {R}}.
\newblock Springer, NY, second edition, 2013.
\newblock http://www.asdar-book.org/.

\bibitem[Broomhead and Lowe(1988)]{Broomhead:Lowe:1988}
David~H. Broomhead and David Lowe.
\newblock {M}ultivariable {F}unctional {I}nterpolation and {A}daptive
  {N}etworks.
\newblock \emph{Complex Systems}, 2:\penalty0 321--355, 1988.

\bibitem[Burrough and McDonnell(1998)]{Burrough:McDonnell:1998}
P.~A. Burrough and R.~A. McDonnell.
\newblock \emph{{P}rinciples of {G}eographical {I}nformation {S}ystems}.
\newblock Oxford University Press, 2nd edition, 1998.

\bibitem[Duchon(1977)]{Duchon:1977}
J.~Duchon.
\newblock \emph{{S}plines minimizing rotation-invariant semi-norms in {S}obolev
  spaces}, pages 85--100.
\newblock Springer-Verlag, Berlin, 1977.

\bibitem[Gong and Flegal(2015)]{Gong:Flegal:2015}
L.~Gong and J.~M. Flegal.
\newblock A practical sequntial stopping rule for high-dimensional markov chain
  monte carlo.
\newblock \emph{Journal of Coputational and Graphical Statistics}, 2015.

\bibitem[Green and Silverman(1994)]{Green:Silver:1994}
P.J. Green and B.~W. Silverman.
\newblock \emph{{N}onparametric {R}egression and {G}eneralized {L}inear
  {M}odels}.
\newblock Champman Hall, London, U.K., 1994.

\bibitem[Gu(2002)]{Gu:2002}
Chong Gu.
\newblock \emph{{S}moothing {S}pline {ANOVA} {M}odels}.
\newblock Springer-Verlag, New York, NY, 2002.

\bibitem[He and Sun(2000)]{He:Sun:2000}
Z.~He and D~Sun.
\newblock {H}ierarchical {B}ayesian estimation of hunting success rates with
  spatial correlations.
\newblock \emph{Biometrics}, 56:\penalty0 360--367, 2000.

\bibitem[Hutchinson and Gessler(1994)]{Hutchinson:Gessler:1994}
M.~F. Hutchinson and F.~R. Gessler.
\newblock {S}plines - {M}ore {T}han {J}ust a {S}mooth {I}nterpolator.
\newblock \emph{Geoderma}, 62:\penalty0 45--67, 1994.

\bibitem[Kimmeldorf and Wahba(1970)]{Kimmeldorf:Wahba:1970}
G.~Kimmeldorf and G.~Wahba.
\newblock {A} {C}orrespondance {B}etween {B}ayesian {E}stimation of
  {S}tochastic {P}rocesses and {S}moothing by {S}plines.
\newblock \emph{Annals of Mathematical Statistics}, 41:\penalty0 495--502,
  1970.

\bibitem[Kimmeldorf and Wahba(1971)]{Kimmeldorf:Wahba:1971}
G.~Kimmeldorf and G.~Wahba.
\newblock {S}ome {R}esults on {T}chebychffan {S}pline {F}unctions.
\newblock \emph{Journal of Mathematical Analysis Applications}, 33:\penalty0
  82--85, 1971.

\bibitem[Kinderman and Monahan(1977)]{Kinderman:Monahan:1977}
A.~J. Kinderman and J.~F. Monahan.
\newblock {C}omputer {G}eneration of {R}andom {V}ariables {U}sing the {R}atio
  of {U}niform {D}eviates.
\newblock \emph{ACM Trans. Math. Softw.}, 3\penalty0 (3):\penalty0 257--260,
  September 1977.
\newblock ISSN 0098-3500.
\newblock \doi{10.1145/355744.355750}.
\newblock URL \url{http://doi.acm.org/10.1145/355744.355750}.

\bibitem[Lang and Brezger(2001)]{Lang:Brezger:2001}
Stefan Lang and Andreas Brezger.
\newblock {Bayesian P-Splines}.
\newblock techreport, University of Munich, 2001.

\bibitem[Laslett(1994)]{Laslett:1994}
G.~M. Laslett.
\newblock {K}riging and {S}plines: {A}n {E}mpirical {C}omparison of {T}heir
  {P}redictive {P}erformance.
\newblock 89:\penalty0 391--400, 1994.

\bibitem[Laslett and McBratney(1990)]{Laslett:McBrat:1990}
G.~M. Laslett and A.B. McBratney.
\newblock {F}urther {C}omparison of {S}patial {M}ethods for {P}redicting {S}oil
  p{H}.
\newblock \emph{Journal of the Soil Science Society of America}, 54:\penalty0
  1553--1558, 1990.

\bibitem[Matheron(1963)]{Matheron:1963}
G.~Matheron.
\newblock {P}rinciples of geostatistics.
\newblock \emph{Economic Geology}, 58:\penalty0 1246--1266, 1963.

\bibitem[Matheron(1973)]{Matheron:1973}
G.~Matheron.
\newblock {T}he intrinsic random functions and their applications.
\newblock \emph{Advances in Applied Probability}, 5:\penalty0 439--468, 1973.

\bibitem[Meinguet(1979)]{Meinguet:1979}
J.~Meinguet.
\newblock {M}ultivariate interpolation of arbitrary points made simple.
\newblock \emph{Journal of Applied Mathematical Physics}, 30:\penalty0
  292--304, 1979.

\bibitem[Nychka(2000)]{Nychka:2000}
D.~W. Nychka.
\newblock \emph{{S}moothing and {R}egression: {A}pproaches, {C}omputation, and
  {A}pplication}, chapter Spatial {P}rocess {E}stimators as {S}moothers, pages
  393--424.
\newblock Chichester, New York, NY, 2000.

\bibitem[Pebesma and Bivand(2005)]{Pebesma:Bivand:2005}
E.~J. Pebesma and R.~S. Bivand.
\newblock {C}lasses and methods for spatial data in {R}.
\newblock \emph{R News}, 5\penalty0 (2):\penalty0
  https://cran.r-project.org/doc/Rnews/., 2005.

\bibitem[Powell(1977)]{Powell:1977}
Michael J.~D. Powell.
\newblock {R}estart procedures for the conjugate gradient methods.
\newblock \emph{Mathematical Programming}, 12\penalty0 (1):\penalty0 241--254,
  1977.

\bibitem[{R Core Team}(2015)]{R}
{R Core Team}.
\newblock \emph{{R}: {A} {L}anguage and {E}nvironment for {S}tatistical
  {C}omputing}.
\newblock R Foundation for Statistical Computing, Vienna, Austria, 2015.
\newblock URL \url{http://www.R-project.org/}.

\bibitem[Rue et~al.(2009)Rue, Martino, and Chopin]{Rue:etal:2009}
Håvard Rue, Sara Martino, and Nicolas Chopin.
\newblock {A}pproximate {B}ayesian inference for latent {G}aussian models by
  using integrated nested {L}aplace approximations.
\newblock \emph{Journal of the Royal Statistical Society: Series B (Statistical
  Methodology)}, 71\penalty0 (2):\penalty0 319--392, 2009.
\newblock ISSN 1467-9868.
\newblock \doi{10.1111/j.1467-9868.2008.00700.x}.
\newblock URL \url{http://dx.doi.org/10.1111/j.1467-9868.2008.00700.x}.

\bibitem[Speckman and Sun({2003})]{Speckman:Sun:2003}
Paul~L. Speckman and Dongchu Sun.
\newblock {F}ully {B}ayesian spline smoothing and intrinsic autoregressive
  priors.
\newblock \emph{Biometrika}, 90:\penalty0 289--302, {2003}.

\bibitem[van~der Linde et~al.(1995)van~der Linde, Witzko, and
  J{\"o}ckel]{vanderLinde:etal:1995}
A.~van~der Linde, K.~H. Witzko, and K.~H. J{\"o}ckel.
\newblock {S}patial-{T}emporal {A}nalysis of {M}ortality {U}sing {S}plines.
\newblock \emph{Biometrics}, 51:\penalty0 1352--1360, 1995.

\bibitem[Wahba(1978)]{Wahba:1978}
G.~Wahba.
\newblock {I}mproper {P}riors, {S}pline {S}moothing and the {P}roblem of
  {G}uarding {A}gainst {M}odel {E}rrors in {R}egression.
\newblock 40:\penalty0 364--372, 1978.

\bibitem[Wahba(1983)]{Wahba:1983}
G.~Wahba.
\newblock {B}ayesian "{C}onfidence" {I}ntervals for the {C}ross-{V}alidated
  {S}moothing {S}pline.
\newblock 45:\penalty0 133--150, 1983.

\bibitem[Wahba(1990)]{Wahba:1990}
G.~Wahba.
\newblock {S}pline {M}odels for {O}bservational {D}ata.
\newblock In \emph{Volume 59 of CBMC-NSF Regional Conference Series in Applied
  Mathematics}. SIAM, Philadelphia, 1990.

\bibitem[Wahba et~al.(1995)Wahba, Wang, Gu, Klein, and Klein]{Wahba:etal:1995}
G.~Wahba, Y.~Wang, C.~Gu, R.~Klein, and B.~E. Klein.
\newblock {S}moothing {S}pline {ANOVA} fpr {E}xponential {F}amilies, with
  {A}pplications to the {W}isconsin {E}pidemiological {S}tudy for {D}iabetic
  {R}etinopathy.
\newblock \emph{Annals of Statistics}, 23:\penalty0 1865--1895, 1995.

\bibitem[Walker et~al.(2011)Walker, Laud, Zantedeschi, and
  Damien]{Walker:etal:2011}
Stephen~G. Walker, Purushottam~W. Laud, Daniel Zantedeschi, and Paul Damien.
\newblock Direct sampling.
\newblock \emph{Journal of Computational and Graphical Statistics}, 20\penalty0
  (3):\penalty0 692--713, 2011.
\newblock ISSN 10618600.
\newblock URL \url{http://www.jstor.org/stable/23248847}.

\bibitem[White(2006)]{White:2006}
Gentry~A. White.
\newblock \emph{Bayesian Semiparametric Spatial and Joint Spatio-Temporal
  Smoothing}.
\newblock PhD thesis, 2006.

\end{thebibliography}

\end{document}